# The Cell Ontology in the age of single-cell omics

Shawn Zheng Kai Tan [1,2,*], Aleix Puig-Barbe [2,*], Damien Goutte-Gattat [3,4], Caroline Eastwood [5], Brian Aevermann [6], Alida Avola [5], James P Balhoff [7], Ismail Ugur Bayindir [5], Jasmine Belfiore [5], Anita Reane Caron [2], David S Fischer [8], Nancy George [9], Benjamin M Gyori [10], Melissa A Haendel [11], Charles Tapley Hoyt [12], Huseyin Kir [5], Tiago Lubiana [13], Nicolas Matentzoglu [14], James A Overton [15], Beverly Peng [16], Bjoern Peters [17], Ellen M Quardokus [18], Patrick L Ray [19], Paola Roncaglia [2], Andrea D Rivera [5], Ray Stefancsik [2], Wei Kheng Teh [2], Sabrina Toro [11], Nicole Vasilevsky [20], Chuan Xu [21], Yun Zhang [22], Richard H Scheuermann [22,^], Christopher J Mungall [23,^], Alexander D Diehl [24,^], David Osumi-Sutherland [5,^]

1. SignaMind, Singapore
2. European Bioinformatics Institute (EMBL-EBI), Wellcome Genome Campus, Hinxton, Cambridge, CB10 1SD, UK
3. Department of Physiology, Development and Neuroscience, University of Cambridge, Downing Street, Cambridge CB2 3DY, UK
4. German BioImaging GMB e.V., c/o University of Konstanz, Box 604, 78454 Konstanz, Germany
5. Wellcome Sanger Institute, Wellcome Genome Campus, Hinxton, Cambridge CB10 1RQ, UK
56. Chan Zuckerberg Initiative
7. Renaissance Computing Institute, University of North Carolina, Chapel Hill, NC, USA
8. Medical University of Vienna, Institute of Artificial Intelligence, Center for Medical Data Science, Vienna, Austria
9. Syngenta, Jealott's Hill, Warfield, Bracknell, UK
10. Laboratory of Systems Pharmacology, Harvard Medical School, Boston, MA, USA
11. University of North Carolina at Chapel Hill, Chapel Hill, NC, USA
12. RWTH Aachen University, Institute of Inorganic Chemistry, Landoltweg 1a, 52074, Aachen, Germany
13. University of São Paulo, São Paulo, Brazil
14. Semanticly, Athens, Greece
15. Knocean Inc., Toronto, Ontario, Canada
16. Department of Informatics, J. Craig Venter Institute, La Jolla, CA, USA
17. La Jolla Institute for Immunology, 9420 Athena Circle, La Jolla, CA 92037, United States
18. Indiana University: Bloomington, Indiana, US
19. Allen Institute for Brain Science, Seattle, WA., United States
20. Critical Path Institute, Tucson, AZ, United States
21. Cambridge Stem Cell Institute and Department of Medicine, University of Cambridge, Cambridge, UK
22. Division of Intramural Research, National Library of Medicine, National Institutes of Health, Bethesda, MD, United States
23. Lawrence Berkeley National Laboratory (LBNL), Berkeley, CA 94720, United States
24. Department of Biomedical Informatics, Jacobs School of Medicine and Biomedical Sciences, University at Buffalo, Buffalo, NY 14203, USA

* These authors contributed equally
^ Correspondence to: do12@sanger.ac.uk (DOS), richard.scheuermann@nih.gov (RHS), cjmungall@lbl.gov (CJM), addiehl@buffalo.edu (ADD).

# Abstract

Single-cell omics technologies have transformed our understanding of cellular diversity by enabling high-resolution profiling of individual cells. However, the unprecedented scale and heterogeneity of these datasets demand robust frameworks for data integration and annotation. The Cell Ontology (CL) has emerged as a pivotal resource for achieving FAIR (Findable, Accessible, Interoperable, and Reusable) data principles by providing standardized, species-agnostic terms for canonical cell types—forming a core component of a wide range of platforms and tools. In this paper, we describe the wide variety of uses of CL in these platforms and tools and detail ongoing work to improve and extend CL content including the addition of transcriptomic types, working closely with major atlasing efforts including the Human Cell Atlas and the Brain Initiative Cell Atlas Network to support their needs. We cover the challenges and future plans for harmonising classical and transcriptomic cell type definitions, integrating markers and using Large Language Models (LLMs) to improve content and efficiency of CL workflows.

# Introduction

An ever growing array of single cell omics techniques are being used to generate massive and diverse datasets profiling single cells. These include techniques for single cell transcriptomics (Nguyen et al. 2018; Shekhar and Menon 2019), proteomics (Bennett et al. 2023), multiplexed imaging of protein location (Hickey et al. 2022), metabolomics (Shrestha 2020), spatially resolved transcriptomics (Wang et al. 2023), connectomics (Winding et al. 2023; Scheffer et al. 2020), epigenetic profiling (Baek and Lee 2020), and multimodal techniques such as patch-sequencing (patch-seq) (Cadwell et al. 2016).

The substantial volumes of data generated by these techniques pose a significant challenge for biologists, clinicians, and bioinformaticians aiming to identify and integrate relevant data for analysis, generate hypotheses, and apply machine learning techniques across datasets. To best utilise these new datasets for hypothesis generation in biology and medicine, we need ways to connect omics data to traditional knowledge of cell types, including their anatomy and physiology. This prior knowledge is primarily found in the research literature as free text in non-standardised forms. Integrating this knowledge with quantifiable 'omics datasets is therefore a challenge.

Findable, Accessible, Interoperable, and Reusable (FAIR) data annotation principles (Wilkinson et al. 2016) can help to solve these problems by ensuring that data is structured and annotated consistently. Annotation with ontology terms is a key component of these principles. In the context of single cell data, ontologies provide unambiguous identifiers (McMurry et al. 2017) for consistent annotation of cell type, tissue, developmental stage, and disease. Referenced textual definitions attached to these identifiers ensure that the meaning of ontology terms is clear, while attached names and synonyms support searching with a wide range of commonly used names.

Ontologies also relate terms to each other in biologically meaningful ways. As a result, ontology annotation not only integrates data annotated with the same term, it can also integrate data annotated with related terms. This underlies the widespread use of the Gene Ontology for gene enrichment analysis (Subramanian et al. 2005; Y. Hu et al. 2023) - summarising the functions of genes in lists by mapping up from specific annotations of individual genes to more general terms in the ontology, e.g. mapping 'CMG complex assembly' to 'lagging strand initiation' to 'DNA replication'. In the context of a cell ontology, a query for data annotated with an ontology term for 'T cell' might use the structure of the ontology to return data annotated with terms for the many known subtypes of 'T cell' (see Figure 3 for other examples).

Relationships between ontology terms can extend beyond a single type of term, for example linking a cell type to terms recording its anatomical location, characteristic components, functions, and cell surface markers (see Figure 1). Collectively, these relationships form a graph that can be used to support applications ranging from database search to gene set enrichment to machine learning and which is used as the backbone of knowledge graphs (Putman et al. 2024; Court et al. 2023). In the context of cell type ontologies, these relationships, combined with referenced textual definitions, provide links to classical and contextual knowledge about cell types. This context is useful not only to biologists but also as input to machine learning (Fischer et al. 2021; Wang et al. 2021).

## The Cell Ontology

The Cell Ontology (CL), the first version of which was released in 2004 (Bard et al. 2005), is a foundational resource for FAIR annotation of cell type data, and widely used to drive data integration and search in biomedical research by academia and their partners, government supported initiatives (Table 2), and industry (eg pharmaceutical companies, tech companies like Benevolent AI, and many others)(Tan et al. 2025). CL represents canonical (non-pathological/wild-type) cell types. It covers granular cell types in mammals, while also supporting general classes for other metazoa (e.g., muscle cell, neuron). It facilitates interoperability through mappings to species-specific ontologies for model organisms such as those for *Xenopus*, Zebrafish, and *Drosophila*.

As part of the OBO (Open Biological and Biomedical Ontology) Foundry (Smith et al. 2007; Jackson et al. 2021), CL adheres to standardized methods for defining relationships and classifications (Diehl et al. 2016; Meehan et al. 2011) and integrates tightly with other OBO Foundry ontologies. CL is also used by phenotype (Gargano et al. 2024), disease (Vasilevsky et al. 2025), and trait ontologies (Stefancsik et al. 2023) to record the cellular location of disease and phenotype and by the Gene Ontology (GO) to record the location of processes and cellular components (Gene Ontology Consortium et al. 2023) - see Figure 1 for examples. CL in turn uses GO to capture cell type functions and the characteristic components of cell types, and the anatomy ontology Uberon (Mungall et al. 2012) to record anatomical location of cell types. As the GO is a rich source of gene product annotation, links between CL and GO provide a potential route to map genes relevant to the transcriptomic and proteomic signatures of cell types.

Terms in CL are predominantly defined and classified based on classical structural, functional, and molecular criteria (classical cell types). For instance, pulmonary alveolar type 2 cells (AT2 cell) (Figure 1) are defined by their location in the alveolar epithelium, their columnar morphology and their function in surfactant production, secretion and homeostasis (Pranty et al. 2025). Surfactant production requires the presence of an alveolar lamellar body, an organelle that is specific to these cells. These properties are all reflected in the CL textual and formal definition - which classifies the cell type as a columnar/cuboidal epithelial cell, records its alveolar location, and links it to GO terms for alveolar lamellar body and surfactant homeostasis (http://purl.obolibrary.org/obo/CL_0002063). As shown in Figure 1 and Table 1, we can find genes that reflect the transcriptomic signatures underpinning the structure and function of this cell type by intersecting the set of genes annotated to these GO terms with markers derived from analysis of transcriptomics data annotated with the AT2 CL term.

CL supports diverse use cases by allowing terms to have more than one parent classification; for example, AT2 cells are classified both as alveolar epithelial cells and cuboidal epithelial cells (other alveolar epithelial cells are squamous). Similarly tracheal goblet cells are both goblet cells (which are found in many mucosal epithelia) and tracheal epithelial cells, reflecting both function and anatomical attributes. Such classifications leverage OWL logic to automate error checking and ensure consistency (Diehl et al. 2016; Jackson et al. 2021; Smith et al. 2007; Meehan et al. 2011).

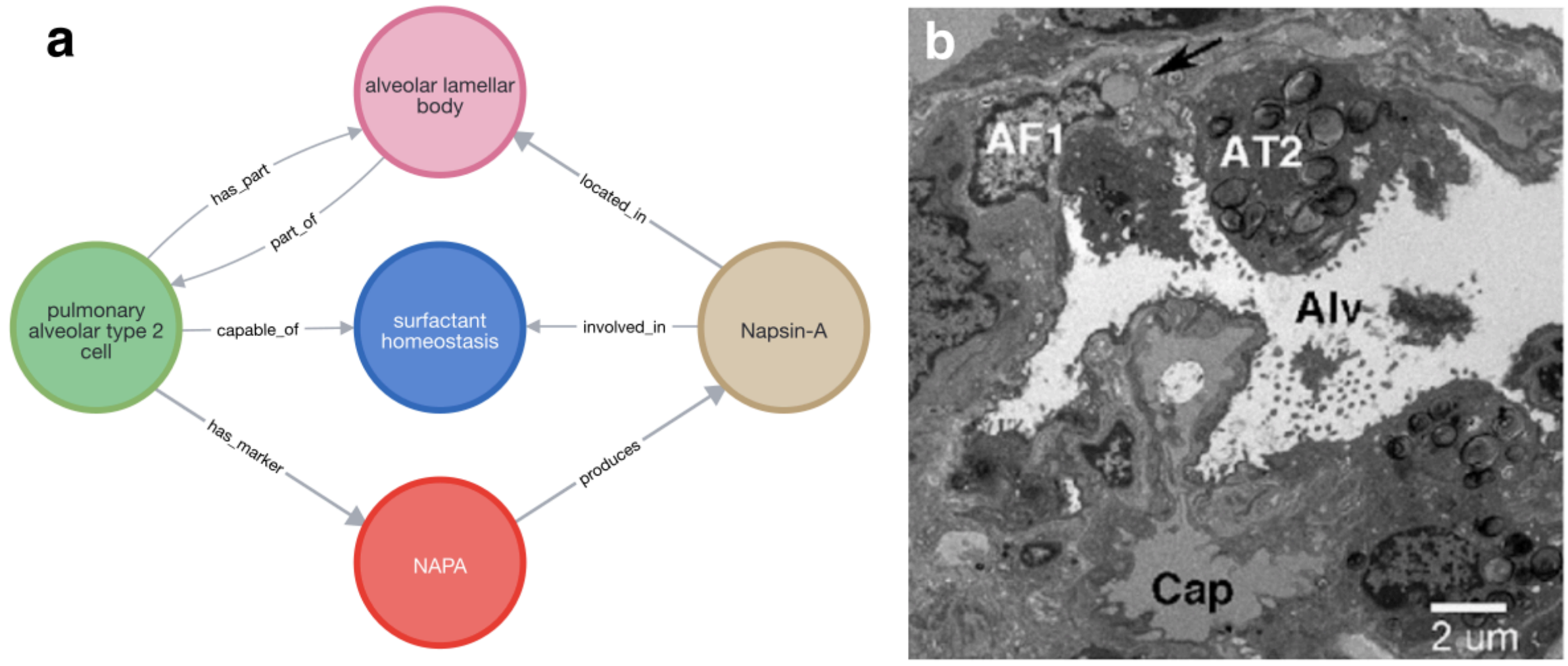

**Figure 1.** *'Classical', property-based definition of pulmonary alveolar type 2 cell (AT2 cell) in the Cell Ontology.* In the Cell Ontology this term (CL:0002063) is defined as "A pulmonary alveolar epithelial cell that modulates the fluid surrounding the alveolar epithelium by secreting and recycling surfactants via specialized organelles called alveolar lamellar bodies… This cuboidal cell is thicker than squamous alveolar cells (AT1), has a rounded apical surface that projects above the level of the surrounding epithelium, and its free surface is covered by short microvilli". **Panel A** shows this property-based definition as a graph linking the Cell Ontology term 'pulmonary alveolar type 2 cell' to the GO terms for 'surfactant homeostasis' and 'alveolar lamellar body'. GO annotation records the function and of the Napsin-A protein in surfactant homeostasis and its location in the alveolar lamellar body. NAPA, the gene encoding Napsin-A, was identified as an AT2 cell marker using by analysis of single cell transcriptomics data annotated with this cell type (CZI Single-Cell Biology Program et al. 2023). **Panel B**, shows an electron micrograph of an AT2 cell with many alveolar lamellar bodies (ALM) (adapted from (Sonder et al. 2023; Sun et al. 2022)).

| Cell Type | GO Term | Mouse Markers | Human Markers |
|---|---|---|---|
| pulmonary alveolar type 2 cell | surfactant homeostasis | Lamp3, Abca3, Lpcat1, Ctsh, Sftpd, Napsa | SFTPD, NAPSA, ABCA3, CTSH |
| pulmonary alveolar type 2 cell | alveolar lamellar body | Abca3, Napsa, Sftpc, Sftpb, Ctsh | ABCA3, NAPSA, SFTPB, SFTPC, CTSH |

**Table 1** shows a table of mouse and human markers of AT2 cells that are annotated to one or both of these GO terms.

Markers are practical tools for identifying cell types. Their practical application depends on both anatomical and experimental context. A gene (or set of genes) is a useful marker if its expression can be used to unambiguously identify (mark) a cell type in some context. Markers are often species-specific. Proteins are commonly used as cell-type markers, but protein and transcript levels often do not correlate well and proteins may be present when transcript has degraded (Vallejo et al. 2022). Markers that are specific in a narrow anatomical context (e.g. a single brain region) may not be in the context of a whole brain (see Figure 2). Many classical markers have their origin in the context of imaging prior to the advent of spatial transcriptomics. In this context, markers typically need to distinguish cell types within the narrow anatomical context of a single imaged tissue slice or block, but large differences in expression compared to background are needed for thresholding to be able to mark cells. In contrast, markers derived from single cell transcriptomics often need to distinguish cells in the context of large and diverse tissue samples, and the typical approach to deriving markers, differential expression analysis, can take advantage of relative expression levels that would be too subtle to clearly mark cells in a traditional imaging context. Variation in documented markers can also come from different marker differential expression algorithms producing different top marker gene lists (Wang et al. 2023).

CL uses gene expression as one criterion for defining cell types in cases where expression is widely considered definitional for the cell type, for instance for immune cells, where cell type definitions have long been dominated by cell surface markers used for isolating cell populations. As detailed in this paper, we also use markers as a component of cell types defined with reference to transcriptomic data. However, CL does not aim to be a comprehensive source of cell type markers. The highly contextual and diverse nature of markers means that storing all markers would bloat the ontology and make it hard to use. A number of resources use CL terms to record markers in their species and anatomical context, including the HubMap Human Reference Atlas ((HuBMAP Consortium 2019; Bueckle et al. 2024), CellMarker (C. Hu et al. 2023) and CELLxGENE Expression (CZI Single-Cell Biology Program et al. 2023). The SOULCAP initiative (Standardised Ontology Unique Labelling for Cytometry Annotation of Populations, https://soulcap.org/) is an international consortium of academic and industry researchers in flow cytometry who are developing an expert curated database of marker-cell type associations, primarily for immune cells, in collaboration with CL developers, as well as bioinformatics methods to standardize cell type identification across flow cytometry results.

**Figure 2.** *The contextual nature of markers*. **Panel A** shows a UMAP embedding of > 4m cells covering the whole mouse brain (Yao et al. 2023), with cells annotated as Cajal Retzius cells (CR) highlighted. Lhx1 is a classically known marker used to identify CR cells in microscopy (Causeret et al. 2021). It is not specific in the context of the whole brain (**Panel B**). However, the vast majority of CR cells surviving in the adult mouse brain are present in the hippocampus (Causeret et al. 2021). When we filter out cells from other brain regions, Lhx1 is a much more specific marker (**Panel C**). This figure was generated using the Allen Brain Cell Atlas (https://portal.brain-map.org/atlases-and-data/bkp/abc-atlas).

## Transcriptomic cell types

Recent advances in single-cell omics technologies have transformed how scientists study cell types by providing unbiased ways of characterising cell identity without relying on prior knowledge. Progress has been made most rapidly in single cell transcriptomics, which has provided new means for typing cells based on shared transcriptomic profiles. This typically involves selection of informative (highly variable) genes, dimensionality reduction, and clustering in latent space (Luecken and Theis 2019). Types defined in this way can be mapped back to classically defined types based on expression of known markers, sometimes combined with gene set enrichment analysis providing evidence for functions/pathways diagnostic of particular cell types (Franchini et al. 2023).

Clusters of sufficient granularity often correspond well to classical cell types or subdivide them (Shekhar et al. 2016; Elmentaite et al. 2021). Where they do not, some researchers merge or split clusters to more closely match classical cell types or align with other modalities, whereas others treat cell types that are transcriptomically defined (T-types) as primary (Zeng 2022). The latter approach is especially important in the central nervous system, where classical cell typing has only ever been able to provide a very limited and incomplete picture of the cell types present (Zeng 2022). As well as defining types, transcriptomic analysis is also used to define hierarchies of cell types based on transcriptomic similarity, including transcriptomic hierarchies covering the entire brain (Yao et al. 2023).

In this paper we detail how CL is currently being used to help make data about cell types Findable, Accessible, Interoperable, and Reusable (FAIR). We explore how CL helps to form a bridge between different data modalities, and between 'omics views of cell types and classical cell type definitions. We also detail how CL is evolving to face new challenges, including extending to include T-types, defined via links to reference data and marker sets derived from reference data. Finally we outline how improved technical and community coordination and the use of large language models in the curation process (Toro et al. 2024; Caufield et al. 2023) is driving content improvements and new opportunities for CL use.

# Results

## Application

### CL drives FAIR across multiple platforms, species and modalities

With the huge influx of single cell 'omics data, annotation with CL has become an important step in making single cell/nucleus RNA-seq (scRNAseq) data FAIR, and has become a key use of CL.

Consortia and platforms like HuBMAP (the Human BioMolecular Atlas Program)(HuBMAP Consortium 2019), the Human Cell Atlas (HCA)(Regev et al. 2017), and CELLxGENE (Megill et al. 2021; CZI Single-Cell Biology Program et al. 2023) mandate the use of CL IDs to annotate cell type terms in datasets—this has allowed better integration and findability of datasets. (For a more comprehensive list of users, see Table 2).

| Project | Description | CL use | Link/Citation |
| --- | --- | --- | --- |
| HuBMAP, Human Reference Atlas | A consortium of research teams that aims to develop a framework for mapping the human body at single-cell resolution. | Maps all its cell types to CL and PCL. | (HuBMAP Consortium 2019) |
| Human Cell Atlas (HCA), Data Coordination Platform | A consortium of researchers that aim to create a comprehensive reference map of every cell type in the human body. | Uses CL to annotate data from cell suspension studies. | (Regev et al. 2017) |
| CZ CELLxGENE | A Web platform (CZ CELLxGENE Discover) that hosts a massive collection of single cell RNAseq data. | Uses CL to annotate cell type and CL structure to drive Findability and Interoperability. | (CZI Single-Cell Biology Program et al. 2023) |
| BGEE | A web platform integrating curated gene expression data across multiple modalities and species. | Uses CL to annotate datasets. | (Bastian et al. 2021) |

| Cell Annotation Platform | A platform for community annotation of scRNAseq datasets. | Uses an autosuggest system to recommend CL terms for annotation leveraging relationships with Uberon tissue terms. | https://celltype.info |
|---|---|---|---|
| Single Cell Expression Atlas | An online bioinformatics resource that enables easy access to information about gene expression across species, tissues, cells, experimental conditions and diseases. | Uses CL and its extensions to annotate scRNAseq data and "anatamogram" schematics illustrating cell type location in organs and tissues. | (George et al. 2024) |
| Brain Data Standards Ontology | A data-driven ontology, with an initial focus on the data generated in the BRAIN Initiative Cell Census Network (BICCN) mini-atlas of the mammalian primary motor cortex. | Uses CL terms as superclasses. | (Tan et al. 2023) |
| CellTypist | A machine learning driven annotation transfer system for annotating cell types in single cell transcriptomic datasets. | All cell types defined by the CellTypist model are linked to CL terms. | (Domínguez Conde et al. 2022) |
| HuBMAP, Azimuth | An online platform that annotates uploaded datasets using annotation transfer from reference datasets. | Uses CL to annotate reference datasets. | https://azimuth.hubmapconsortium.org/ |
| Ontology-based single cell Classification (OnClass) | An algorithm and accompanying software for automatically classifying cells from scRNAseq experiments. | Uses CL graph to infer cell type relationships. | (Wang et al. 2021) |
| Sfaira | A single-cell data zoo for public data sets paired with a model zoo for executable pre-trained models. | Uses CL to annotate cell type terms and the ontology graph to train a supervised model for cell type identification. | (Fischer et al. 2021) |
| Cell BLAST | A cell-querying method built on a neural network-based generative model and a customized cell-to-cell similarity metric that aims to be a one-stop solution for real-world scRNA-seq cell querying and annotation. | Uses CL to annotate cell types and ontology structure for improved assignment. | (Cao et al. 2020) |

| CellMarker 2.0 | A database of Cell Markers. | Uses CL and Uberon to record cell type markers and their tissue context in human and mouse. Data is curated from the literature and supplemented by integrated datasets and data analysis tools. | (C. Hu et al. 2023) |
|---|---|---|---|
| Cell Image Library | An online resource of images of cells, their substructures and phenotypes. | Uses CL in annotation of microscopy images. | (Orloff et al. 2013) |
| human Ensemble Cell Atlas (hECA) | A unified informatics framework for seamless cell-centric data assembly. | Maps names to CL terms. | (Chen et al. 2022) |
| celldex | An R package containing reference datasets with cell type information (“Pokédex for Cell Types”). | Maps names to CL terms. | https://bioconductor.org/packages/devel/data/experiment/vignettes/celldex/inst/doc/userguide.html |
| Cellarium | A machine learning driven annotation transfer system that leverages the CELLxGENE scRNAseq corpus and its ontology annotation. | Leverages CL annotation on CELLxGENE. Outputs annotation with CL terms. | https://cellarium.ai/tool/cellarium-cell-annotation-service-cas/ |
| popV | Annotation transfer with Confidence scores. Polls the results of multiple annotation transfer algorithms including OnClass. | Uses CL as both input and output. | (Ergen et al. 2024) |

**Table 2.** A non-exhaustive list of resources that use the Cell Ontology.

This is perhaps best illustrated by its use in the CZ CELLxGENE (CxG) platform (CZI Single-Cell Biology Program et al. 2023), the largest and most widely used, consistently annotated database of single cell transcriptomics data. It uses CL to annotate single cell transcriptomics data and to drive many aspects of its functionality.

CxG mandates that CL be used for annotation of cell types in all submitted datasets via a well-defined standard for submission of annotated cell-by-gene matrices (CZI Single-Cell Biology Program et al. 2023). By providing terms for annotation at various levels of granularity, it allows for accurate but less specific annotations where annotators are not confident of the precise identity of the cells being annotated. On the CxG Discover platform, users can find datasets by the cell types they contain via a faceted browsing interface, which supports searching for cell types by name and browsing via a simplified version of CL hierarchy (Figure 3C). An alternative view is provided by CxG CellGuide, which displays extended information and analysis about cell types and links to datasets featuring data about these cell types. CellGuide supports search for cell type and dataset by cell type name or synonym (from CL) and browsing of cell types and datasets via a more complete, interactive graph view of CL hierarchy (Figure 3A&B). CellGuide also takes advantage of CL annotation and hierarchy to calculate marker genes by aggregating data about cell types across multiple datasets and multiple levels of granularity (Figure 3D). CxG also provides integrated embeddings, combining data about specific cell types from across the CxG corpus. For example, the integrated embedding page for human T cells (https://cellxgene.cziscience.com/e/cellguide-cxgs/homo_sapiens/CL_0000084.cxg/) aggregates data annotated with over 60 terms for types of T cells from CL. CL annotation also plays important roles in CxG Census API functionality, supporting the generation of synthetic cell-by-gene matrices for specific cell types from across the census. In summary, CL is an integral component of CELLxGENE that is critical to much of its functionality.

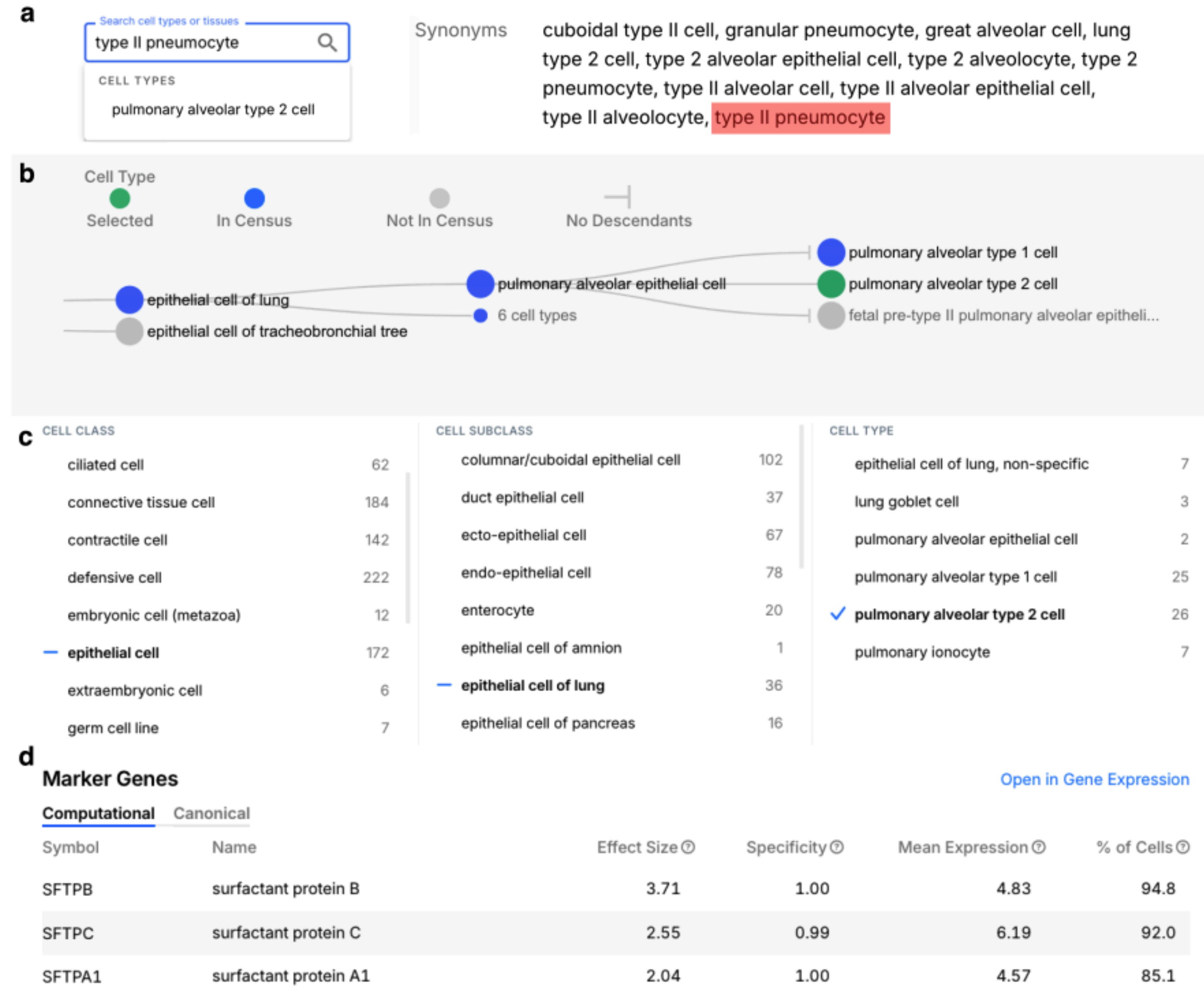

**Figure 3.** The Cell Ontology supports findability of scRNAseq data on the CZ CELLxGENE platform (CZI Single-Cell Biology Program et al. 2023). CELLxGENE CellGuide (panels A and B) provides data

about markers and other summary information about cell types as well as links to datasets. **Panel A:** CellGuide Search supports search by synonym, in this case the term 'pulmonary alveolar type 2 cell' is found via its synonym ‘type II pneumocyte’. **Panel B**: CellGuide supports browsing of related terms via a compact, interactive tree that shows the classification hierarchy of CL, annotated to show which terms have associated data. Selecting a term returns datasets annotated with that term and its subclasses. **Panel C** shows a search for datasets by cell type using the "CELLxGENE Discover" search interface. An initial lexical search step narrows down search terms from all those used. For ease of browsing the results are arranged in a 3 level hierarchy that combines curation of terms into 3 lists of terms at different levels of granularity ("class", "subclass", "type") with links between the levels driven by the structure of CL. In the above example the 'Class' level term selected is 'epithelial cell', which restricts the 'Subclass' column terms to subclasses of 'epithelial cell' in CL including the selected term 'lung epithelial cell' of which the selected "Cell Type" term, ‘pulmonary alveolar type 2 cell’ has been used in 15 of >1000 datasets on CELLxGENE. **Panel D:** Annotation with CL terms (in this case pulmonary alveolar type 2 cell) is used to aggregate single cell transcriptomic data to calculate marker genes.

CL is also used, along with Uberon, by the HuBMAP Human Reference Atlas (HuBMAP Consortium 2019; Börner et al. 2025) and CellMarker (C. Hu et al. 2023) to curate sets of cell-type specific markers in various tissue contexts from the literature. This allows the integration of community curated markers from the HuBMAP Human Reference Atlas into multiple resources including the CxG CellGuide and the gene enrichment knowledge graph Enrichr-KG (Evangelista et al. 2023).

While the primary focus of CL is on vertebrate and especially mammalian cell types, precise description of specialized non-mammalian cell types is delegated to species-specific ontologies. CL provides “bridges” between its general cell type terms to their species-specific counterparts. These bridges are routinely curated to ensure their continued accuracy, and the workflow used to produce them has recently been modernized (as further described in the Methods section). This supports integration of cell types specific to a much broader range of species including Zebrafish (Van Slyke et al. 2014), *Xenopus* (Segerdell et al. 2013), *Drosophila* (Costa et al. 2013; Court et al. 2023), and *C.elegans* (Davis et al. 2022). While some of these bridges are likely to represent convergent phenotypes, especially between vertebrates and invertebrates, recent transcriptomic evidence supports conservation of broad cell types (e.g. neurons and muscle cells) across the bilateria (Tarashansky et al. 2021).

These cross-species bridges do not explicitly encode whether related cell types are homologous or analogous. However, single-cell transcriptomics is enabling more systematic evaluation of cross-species relationships. Where transcriptomic analyses provide strong evidence for orthology, we are beginning to add grouping terms that unify cell types across species, as discussed further in the Discussion section.

CL support for integration across diverse species is well illustrated by the EBI Single Cell Expression Atlas (George et al. 2024). For example, a search for enterocyte, for which evidence exists for conservation between vertebrates and invertebrates (Ohlstein and Spradling 2006), finds data annotated with specific types of enterocyte in *Drosophila*, humans and mice (Figure 4).

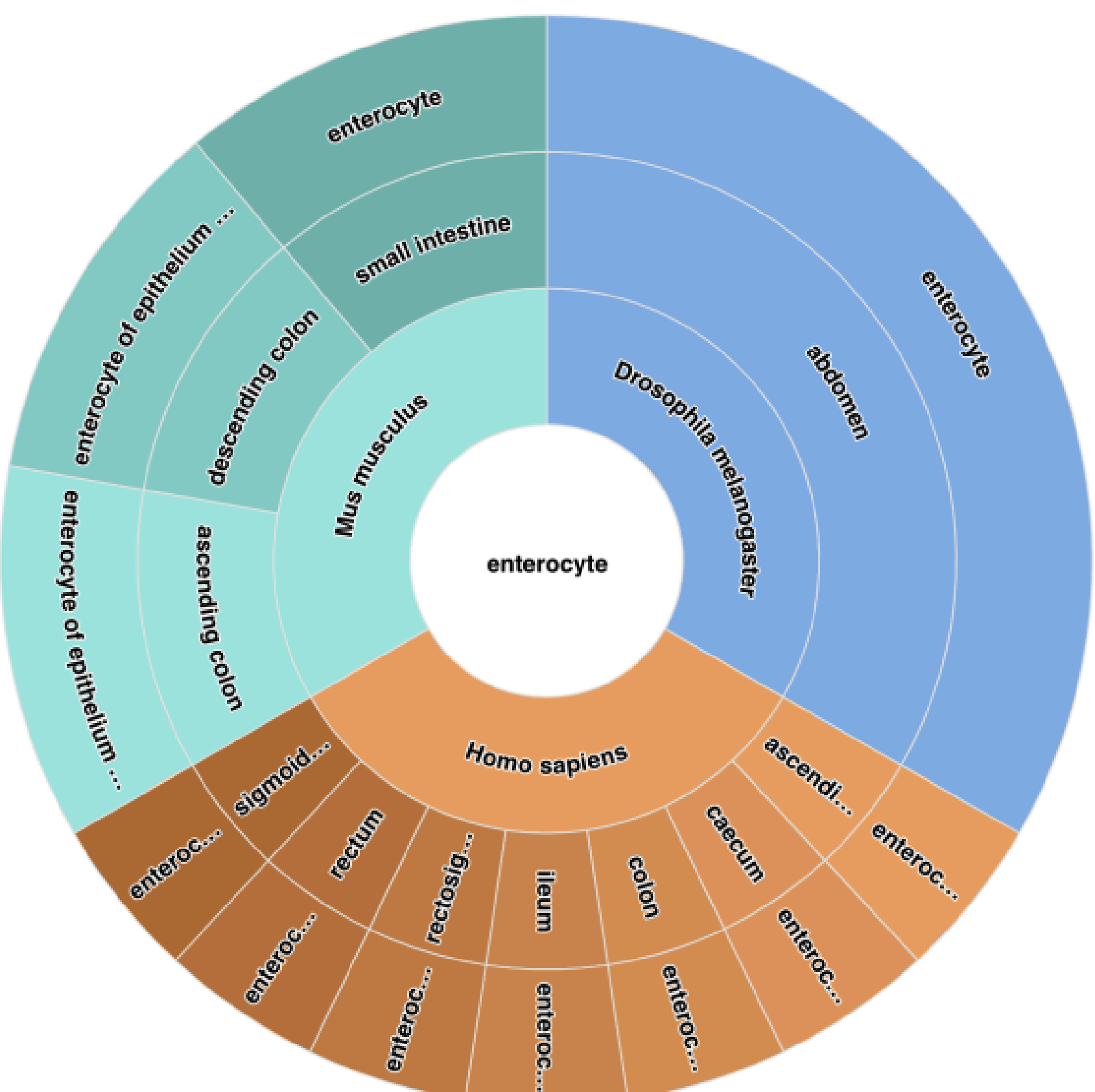

**Figure 4.** Use of extended CL for cross-species data integration on the EBI single cell expression atlas. The central circle shows a general cell type (in this case enterocyte) found by lexical search on the atlas. The inner and middle wheel show the species and tissue of origin of the sample used as input to scRNAseq. The outer wheel displays subtypes of enterocyte used to annotate single cell transcriptomics data. Clicking on the outer wheel takes users to a list of highly expressed genes in the cell type and to individual studies. This search is made possible by an extended version of CL that folds in cell type terms from other, single-species ontologies (in the above example, the Drosophila Anatomy Ontology) via curated bridges between the two ontologies using SSSOM (see Methods) (Matentzoglu, Balhoff, et al. 2022).

While much recent usage of CL has focussed on annotating transcriptomics data, one great strength of CL and the ontologies that extend it is their utility for annotating data across modalities. This can perhaps be best seen in the use of the Drosophila Anatomy Ontology to annotate 3D images (>100,000 images from confocal and EM data), connectomics (>4m pairwise connections) and transcriptomics data (240 cell types, 1560 annotations, 2.5m gene expression links) using cell type terms that extend CL (Court et al. 2023). CL is used directly to annotate microscopy images and support their Findability on the Cell Image Library (Orloff et al. 2013) and to annotate schematic images of cell types in anatomical context as part of the Single Cell Expression Atlas (George et al. 2024).

## CL use in Automated Cell Type Annotation

Manual annotation of scRNAseq data is slow and labour intensive. In tissues for which high-quality annotated reference data is available, automated annotation is critical to efficiently interpreting scRNAseq data in the context of disease and experimental manipulation (Hu et al. 2025). CL is increasingly integrated into tools for automated cell type annotation including the Azimuth annotation transfer tool (Hao et al. 2021), CellTypist (Domínguez Conde et al. 2022; Xu et al. 2023), and scTAB (Fischer et al. 2024) which both feature mappings to CL. It is also used directly by multiple automated annotation tools as a source of knowledge to help drive annotation.

OnClass is an automated annotation system, built on annotated training data (Wang et al. 2021), that uses CL as an input. Typically such systems are limited to annotating cell types that are present and annotated in their training data. Onclass takes advantage of the observation that relatedness in the ontology graph correlates with transcriptomic similarity, along with similar correlations with text similarity of labels and definitions, to support annotation of cell types in transcriptomic data that are present in CL but not in the training data. The same team has further extended their work in a successor software that uses 8 different annotation methods, including OnClass, and CL, to create an ensemble of prediction models with a voting scheme (Ergen et al. 2024).

A similar approach is taken by CellO (Bernstein et al. 2021), a discriminative, supervised machine learning approach for classifying clusters of cells in human scRNA-seq data. CellO uses the structure of CL combined with extensive training data to support automated hierarchical annotation with CL terms.

Sfaira (Fischer et al. 2021) is Python software designed to enable scRNA-seq data curation, ontology-based database queries and automated analysis with pre-trained neural networks using data from a variety of curated sources. It takes advantage of the structure of CL beyond the definition of sets of mutually exclusive cell types. Sfaira requires only that annotated cell types be related via a directed acyclic graph and so can take advantage of the multi-inheritance classifications hierarchies in CL. Sfaira also allows for ontology-aware subsetting queries, for example allowing for the selection of any T cell or subtype thereof, to build cross-study meta analyses.

Given that many standards like CELLxGENE (Megill et al. 2021) require the use of CL for cell type annotation, by providing mappings to CL, these tools greatly aid in preparing data for submission.

## Applications for viewing and querying the Cell Ontology

The content of CL can be viewed on a variety of ontology portals (Bioportal (Noy et al. 2009; “Cell Ontology in Bioportal,” n.d.)), Ontology Lookup Service (Jupp et al. 2015; McLaughlin et al. 2025; “Cell Ontology in Ontology Lookup Service (OLS),” n.d.)) and queried via their APIs. More powerful query options are supported by UberGraph.

As already described, CL is tightly integrated with other OBO Foundry ontologies, which makes powerful new types of queries possible, leveraging the knowledge encoded in multiple linked ontologies. For example, we can take advantage of the thousands of asserted and inferred links from Uberon to CL to query for all cell types found in a specific tissue. These queries are exposed through the Ubergraph (Balhoff et al. 2022), which integrates a large and growing set of OBO Foundry ontologies and precomputes inferred relationships. The result is queryable via a SPARQL endpoint and a REST-API that wraps precomposed SPARQL queries. Wrapped queries include a query for cell type by location.

# Community

In keeping with the ethos of the OBO Foundry (Jackson et al. 2021), CL is built as an open, community-driven resource where the democratization of, contribution to, and editing of the ontology is a key consideration. This approach supports distributed maintenance and extension by a team of editors, funded by and engaging with multiple projects and organizations with an interest in its development, including the BRAIN Initiative Cell Atlas Network (BICAN) (Tan et al. 2023), HuBMAP (HuBMAP Consortium 2019; Börner et al. 2021), the HCA (Regev et al. 2017), Cell Annotation Platform (CAP; https://celltype.info/), and the Chan Zuckerberg Initiative (CZI). As a result of this we have seen a large increase in engagement as measured by increased GitHub requests and substantial improvements in responsiveness since 2018 (see supplementary figure 1 for details).

The HuBMAP Human Reference Atlas (HuBMAP Consortium 2019) is a major source of expert input to CL content. Groups of subject matter experts work with HuBMAP curators and developers to curate hierarchies of anatomical terms, cell types and markers covering human anatomy, mapping these terms to CL and Uberon. These efforts are a major source of expert feedback on missing terms, missing relationships, term naming, and term definitions. Additionally, validating these curated hierarchies against both asserted and inferred relationships in CL and Uberon provides feedback and correction for both CL and the curated hierarchies (Caron et al. 2024).

CL editors also work closely with the Human Cell Atlas data-integration efforts to develop integrated atlases for specific organs and tissues, combining multiple independently generated datasets. One early success of this effort is an integrated atlas of the human respiratory system (Sikkema et al. 2022). Ongoing efforts are centered around annotation of integrated reference datasets on the Cell Annotation Platform (CAP) including gut, eye, liver, lung, kidney, and breast. This collaboration has created a productive feedback cycle: CL editors manually review annotations from CAP-hosted atlases, using publicly available spreadsheets to suggest more granular existing CL terms or flag missing ones that are then added to CL to improve coverage across tissues. This hands-on process directly improves CAP annotations and expands the ontology. For example, while reviewing the trabecular meshwork dataset on CAP, we identified gaps and subsequently added missing cell types such as “Schlemm’s canal endothelial cell”, “juxtacanalicular tissue cell”, and “beam cells”.

In a similarly productive cycle with Brain Initiative Atlases, CL editors have utilised LLM Deep Search methods (Xu and Peng 2025; Skarlinski et al. 2024) with information from transcriptomic type annotations (markers, neurotransmitters, location, neuropeptide) to search the literature to identify potential mappings between transcriptomic and classical types along with data and criteria that can be used to test those mappings. In an increasing number of cases, this supports both improved atlas annotation and the definition of terms in CL that unify transcriptomic and classical definitions. For example, we used Deep Search to identify papers and datasets that bridge classical interneuron types and transcriptomic types in the mouse Cerebellum. We then used annotation transfer to identify the same types in the BICAN mouse whole brain atlas, validating results against location from spatial transcriptomics. As a result we were able to improve annotation of the mouse whole brain atlas and types for mouse Lugaro and Candelabrum interneurons that bridge classical and transcriptomic definitions (Osorno et al. 2022; Kozareva et al. 2021; Rivera and Osumi-Sutherland 2025).

The Cell Ontology increasingly provides a bridge between these atlas efforts by defining and linking terms for common cell types found across these tissues. For example, CL has a general term for 'goblet cell', which classifies terms for more specialised goblet cells present in the gut, lung and eye. In parallel, an ongoing collaboration aims to mine CZI datasets (covering a much wider range of datasets than those covered by HCA and BICAN) for new terms and to improve definitions of existing terms, with input from CZI data curators and experts as part of CZI organised expert jamborees.

The BICAN-funded Human and Mammalian Brain Atlas (HMBA) is developing atlases of transcriptomic cell types across mice, humans and non-human primate brains, integrating these atlases across species. Extensive spatial transcriptomics and patch-seq data are being used to map the cell types defined in these atlases to brain regions, functions, and morphology. CL is working with the community developing these atlases to extend CL with classically defined cell types and, as consensus atlases become sufficiently stable, with data-linked transcriptomic cell types.

## Supporting community input to CL

Building high-quality biomedical ontologies often requires close collaboration between expert biologists, who may not have the technical expertise needed for editing ontologies, and ontology editors/engineers, who may not be as well-versed in specific biological knowledge. CL has taken a number of approaches to help bridge this gap.

A range of GitHub issue templates guide users to provide the required information for various types of edit, including addition of new terms and synonyms and edits to existing definitions and relationships. For example, when requesting a new term, we now ask for a preferred term label, synonyms with references, a definition with reference, the anatomical structure the cell is part of with instructions to use UBERON terms, the submitter’s ORCID, and any notes or concerns. Submitters providing ORCIDs get attribution on edited terms, ensuring due credit and fostering better community relations.

Alternatively, users can request batches of related terms using a spreadsheet based standard: Minimal Information Reporting About a CelL (MIRACL)(Lubiana et al. 2022). This has been successfully used to facilitate the addition of sets of terms specified by experts and curators from the HuBMAP Human Reference Atlas (HuBMAP Consortium 2019; Börner et al. 2021, 2025). A Google Sheets plugin is available to support this (Bayindir 2025).

CL also supports the use of Knowledge Graph Change Language (KGCL) (Hegde et al. 2024) to facilitate direct editing of CL by non-experts. KGCL describes ontology editing operations (e.g. renaming a class, changing its parent, adding a synonym, etc.) using a syntax that is very close to natural English language. We allow the use of KGCL instructions on our issue tracker, so that users can directly specify edits in tickets, for example including a KGCL instruction such as “rename CL:0002075 to ‘brush cell of tracheobronchial tree’”. The specified edits are automatically implemented (via GitHub Actions) and are instantly available for review by submitters and editors (via GitHub Pull Requests).

The Biomappings project (Hoyt et al. 2023) enables community-driven curation of high quality, precise ontology mappings using the Simple Standard for Sharing Ontological Mappings (SSSOM) (Matentzoglu, Balhoff, et al. 2022). CL imports these using the ODK plugin for SSSOM, resulting in the addition of 81 mappings to the Medical Subject Headings (MeSH) and 74 mappings to Brenda Tissue Ontology (BTO) at the time of writing. Relatedly, CL has used the Bioregistry (Hoyt et al. 2022) to standardize references to external ontologies and vocabularies in its curated database cross-references.

Finally, we facilitate community review of proposed changes to CL (on GitHub Pull Requests) by displaying human readable summaries of changes to the ontology and their implications for automated classification of cell types. This makes review by editors and contributing experts easier, aided by GitHub social media functionality.

### Improvements to quality control and release frequency

Recent technological improvements have made the process of editing and maintaining CL more efficient and consistent. By adopting tools like the Ontology Development Kit (ODK) (Matentzoglu, Goutte-Gattat, et al. 2022), the editorial team can release updates more frequently—around once a month compared to one or two releases per year before. The ODK also supports automated quality checks that catch common technical issues before they are incorporated and ensure that release artifacts are compliant with the OBO Foundry principles (Jackson et al. 2021). Additionally, CL ensures its compatibility with other related ontologies through advanced tools like Ubergraph, which conducts daily checks for cross-ontology consistency (Balhoff and Curtis 2021).

## Content

The Cell Ontology has significantly expanded over the past few years, with a total of 3228 terms and 3205 synonyms (release v2025-12-17) based on 1448 unique references. Since 2019, we have added over 1000 new cell type terms, and 1051 synonyms. Over 1100 definitions have been added or revised.

While many of these improvements to content have been driven by the needs of single cell transcriptomics atlasing projects, the majority of this work to date has focused on adding classically defined cell types—specifying a general classification and a set of properties which distinguish that cell type and are practically useful in its identification. These definitions, as well as changes to names and the collection of synonyms, reflect evolving domain knowledge and terminology from various disciplines. This is sufficient to support annotation of single cell transcriptomics data wherever annotated cell sets map cleanly to classically-defined cell types (subject to consensus agreement, backed by multi-modal marker evidence) or when further investigation of cell types defined by single cell transcriptomics determines properties that can be used for new classical cell-type definitions in CL. An alternative approach, trialled in the Provisional Cell Ontology (Tan et al. 2023), extends CL with cell types and classifications defined by reference single cell transcriptomics atlases. This approach is especially important for tissues and where classical cell typing has been overwhelmed by complexity and scale of cell types and tissue structure. This is a major issue for neuron types in the brain and CNS cell types (Keefe and Nowakowski 2019; Zeng 2022). It may also be key where transcriptomics is uncovering a wealth of previously uncharacterised diversity, for example, among fibroblasts (Brügger and Basler 2023) and amacrine cells of the retina (Li et al. 2023).

CL also remains committed to representing vetted and standards-compliant cell types defined by various communities of biologists. This work runs in parallel with work to associate findings of scRNAseq and other omics approaches with existing cell types and identify novel cell types. In this section we discuss some of the key improvements in the content of CL.

## Immune refactoring

Significant prior work has gone into representing immune cells of various lineages in CL (Masci et al. 2009; Diehl et al. 2011; Meehan et al. 2011), and it remains important for CL to stay up-to-date with cell type names that align with standard nomenclature used by immunologists. For example, innate lymphoid cells (ILC) are a group of related lymphocytes separate from B cells and T cells that have become widely studied in the past 15 years (Artis and Spits 2015; Ebbo et al. 2017). ILCs lack the adaptive immune receptors of T cells and B cells and rely on combinations of other receptors for activation and inhibition. ILCs are divided into three main groups, the ILC1, ILC2, and ILC3, based on functional and marker-based criteria (Artis and Spits 2015; Spits et al. 2013). Cell types within one ILC lineage, the natural killer cells, have been extensively studied over a much longer period and have been retrospectively grouped with the larger set of ILC1 cells. In collaboration with the Human Immunology Project Consortium and subject matter experts (Dr. Lewis L. Lanier), CL curators enriched CL with a full representation of these cell types, including both generic and human-specific forms axiomatized with their distinguishing surface markers and formal statements of species specificity. This resulted in the addition of new cell types, the revision of existing cell types, and the revision of the lymphocyte hierarchy to properly accommodate the ILC subgroup.

## Unifying the representation of classical cell types across tissues

CL editors collaborate with experts in the HCA integration networks, HuBMAP and community meetings organised by CELLxGENE to ensure that the representation, naming and classification of classical cell types matches the needs of these atlases as closely as possible and is integrated into broader, cross tissue classifications.

This work includes ensuring term labels and synonyms correspond to current community usage (e.g. replacing the outdated label 'pneumocyte' with the more widely used 'pulmonary alveolar epithelial cell'), or to new standards emerging from integration efforts. Editors improve cell type classifications and definitions, aligning them across tissues, enhancing links to the Gene Ontology, and improving the accuracy of anatomical location links. Efficient improvement of textual definitions is supported by the use of Large Language Models (LLM) and standardised prompts, which provide first-pass textual descriptions with links to the literature. Editors then check the assertions in these descriptions for accuracy and relevance, refining the definition. We also make use of OntoGPT (Caufield et al. 2023) to mine the text for candidate links to Gene Ontology terms.

Textual definitions in CL are increasingly supplemented by longer, LLM generated, expert reviewed descriptions from CellGuide. These are further reviewed for inclusion in CL by editors, and only rejected where there are clear errors or where very general term definitions are simply enumeration of subclasses. To date, 434 out of the 607 reviewed Cellguide definitions (71%) have been accepted and integrated into CL, with the remaining 183 definitions awaiting future review and integration. While definition content needs to be relatively minimal, species-general, and apply to all subclasses, these descriptions have no such restrictions. They can therefore include broad, medically relevant context that would not be appropriate in a core definition. This is made clear by additional text attached to descriptions.

LLMs also help editors search for openly licensed figures to incorporate into Uberon and the Cell Ontology. Schematic figures that illustrate the location and anatomical context of cell types are particularly valuable for clarifying the meaning of terms to both editors and users. These figures are linked to terms via a standard schema supported by ontology portals including the Ontology Lookup Service, which makes linked figures visible on term pages.

A major focus of recent work has been on mucosal epithelial cells across many tissues including brush border cells, found in the choroid plexus, kidney tubules, and gut (enterocytes); goblet cells found in multiple mucosa, multiciliated epithelial cells found in many organs (Chae et al. 2023) and tuft cells. Work on tuft cells (also known as “brush cells”) across various epithelial tissues and organs—including those in the gut, respiratory tract, gallbladder, thymus, pancreas, salivary gland and urethra—illustrates this approach well. Tuft cells are characterised by a distinctive “tuft” on their apical surface, allowing them to serve as chemosensors and modulate immune responses across tissues (Schneider et al. 2019). Through recent edits, we have transformed an incomplete set of terms that previously lacked key cell types, had missing links to GO, and only connected to general anatomical locations into a unified hierarchy of 23 tuft cell terms across 10 different organs. We added 12 new CL terms and revised existing terms within the hierarchy—ensuring each term has a detailed definition, appropriate links to GO terms (tuft, cytospinule), and a part of the relationship to a specific epithelium. Accompanying work in Uberon has ensured that the epithelia themselves are represented and classified correctly.

Our improvements highlight the unique characteristics and diverse roles of tuft cells across different tissues. In the thymus, tuft cells are distinguished by their lateral microvilli structure and MHC class II expression (Miller et al. 2018). In recent edits, we have added this term and linked it to a richly-annotated GO process term ‘antigen processing and presentation of endogenous peptide antigen via MHC class II', highlighting its role in central tolerance. In the intestine, we have expanded developmental relationships, linking intestinal tuft cells to intestinal crypt stem cells, reflecting their potential as reserve stem cells after damage (Gerbe et al. 2012). In the nasal cavity, 'tuft cell of olfactory epithelium' (Ualiyeva et al. 2024), has a definition leveraging both single cell transcriptomics data and microscopy data using markers uncovered by transcriptomics, and knockout phenotypes. The resulting definition includes molecular, functional properties, and location, as well as recording the terms used in scRNAseq annotation as synonyms. These changes collectively set a precedent for our future work on unifying the representation of cell types across tissues in CL.

To streamline the process of adding terms to CL, generating textual definitions for tuft cells across the hierarchy was supported by the use of LLMs with standardised prompts. A manual editing step was included to validate the provided literature and refine the definitions as needed. An example of thymic tuft cells is illustrated in Figure 5.

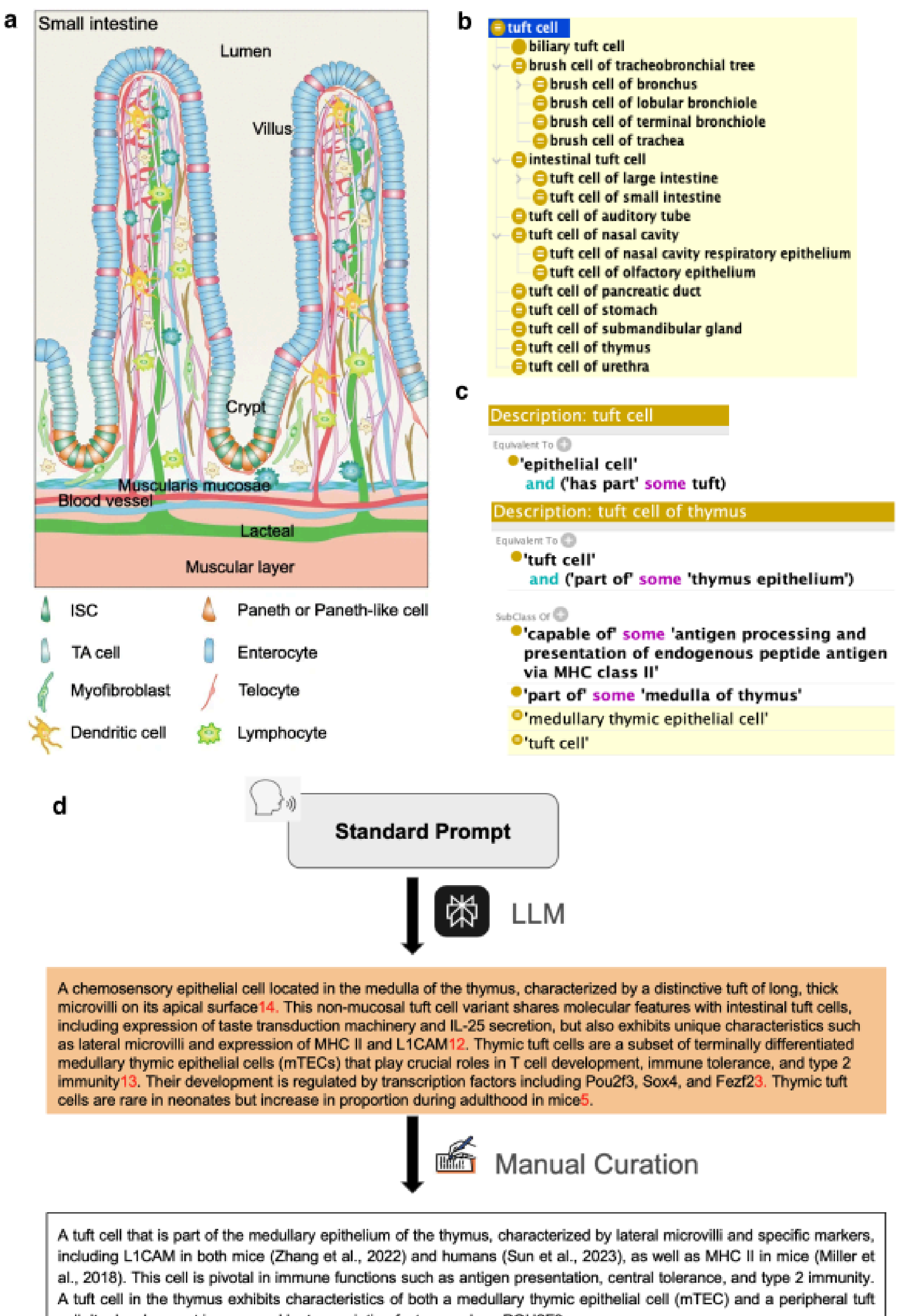

**Figure 5. Panel A**: A schematic of cell types in the intestinal villi of the small intestine (Zhu et al. 2021), represented in CL and linked to gut cell type terms within the ontology. Many of these cell types are specific examples of more general types found across tissues, e.g. tuft cell and goblet cell. Edits to CL have focused on standardising the naming, representation and classification of these cell types across tissues—linking the terms to specific epithelial locations and GO terms for parts and functions. **Panel B**: A unified hierarchy of tuft (bush) cells. **Panel C**: Logical definition of the term 'tuft cell' and of one of its subclasses, 'tuft cell of thymus', following a standard design pattern. **Panel D**:

An example for generating and refining cell type definitions in CL using large language models, illustrated by the tuft cell of thymus.

This process of refactoring and extending CL has been driven by the implementation of unified design patterns for terms. One critical aspect has been unifying the general cell type (genus) terms used in logical definitions, which is critical to ensure that classification can be automated. For example, in earlier version of CL, many epithelial cell types, including multiciliated cells, were manually classified using a mix of subtypes of epithelial cell differentiated by origin (e.g. endo-epithelium), morphology (e.g. columnar), anatomical location or the presence of specialized cell components. As we work on improving epithelial cell type definitions for specific tissues or cross tissue type, we switch to classifying with a general epithelial cell term and classifying dynamically by the properties asserted for individual cell types using an OWL reasoner. Figure 5 shows some example patterns for multiciliated cells.

To enable correct automated classification with a reasoner, sometimes an overhaul of more generalised classification is needed. For example, CL previously classified endothelial cells as a type of epithelial cell. While this is a defensible position, the majority of the biological literature and working biologists treat epithelial and endothelial cells as distinct and non-overlapping categories. By switching CL to this view we are now able to dynamically classify epithelial cell types without automatically classifying endothelial cells under them. This work resulted in automated reclassification of over 40 cell types.

## Extending the Cell Ontology with T-types

Single cell transcriptomics data provides an opportunity to root cell type definition and classification in unbiased analysis of data covering near-complete profiles of gene expression across large populations of individual cells. CL is therefore committed to incorporating T-types—cell types largely defined by their shared transcriptional profiles. CL incorporates T-types based on an interconnected set of evidence criteria. First, we prioritize multimodal evidence beyond transcriptomics alone, incorporating anatomical location, morphology, or functional properties wherever possible. Second, we require that T-types be defined in an integrated sc/snRNAseq atlas from multiple sources or be supported by multiple independent atlases. Third, we require community agreement on reference datasets, typically mediated by official organizations such as the Human Cell Atlas, LungMap, or the Brain Initiative Cell Atlas Network (BICAN). Importantly, we treat all mappings as hypotheses and provide explicit documentation of supporting evidence, allowing users to assess reference quality for themselves.

CL terms require clear definitions that distinguish the cell types they denote from other cell types and can guide accurate annotation of data. Before we can generate definitions for T-types, we need sufficient community consensus on which reference data and analysis to use. Defining a T-type with reference to multiple, independently generated datasets is challenging. While independently generated datasets may be mapped to each other via annotation transfer following statistical matching (Hu et al. 2025), alignment is often imperfect due to experimental and computational differences. These include differences in anatomical regions used as input, e.g. whole cortex vs cortical region (Jorstad et al. 2023), single cell vs single nucleus (Ding et al. 2020), library preparation and sequencing differences. And some annotation transfer tools, like Azimuth and CellTypist, force annotations of novel cell types onto related cell types present in their reference (Hu et al. 2025). Clustering can differ due to study-specific choices of parameters for dimensionality reduction, nearest neighbor search, community detection (Louvain or Leiden), nonlinear embedding (tSNE or UMAP), and iterative/hierarchical clustering scheme (different resolution in each iteration) (Kiselev et al. 2019).

Given this variation, we rely on community agreement on reference datasets, typically a single annotated dataset or integration of multiple datasets sanctioned via an official organization such as the Human Cell Atlas, LungMap, or the Brain Initiative Cell Atlas Network (BICAN) and released with cleanly referenceable versions.

Once we have a reference dataset, we define T-types via links to reference data (in versioned reference datasets) and via marker sets derived from this reference data, folding in additional classical properties where these map clearly, as illustrated in Figure 6. Links to data follow a formal schema established in prior work on the Provisional Cell Ontology (Tan et al. 2023; Bakken et al. 2017)). Marker sets are recorded along with key contextual information including species and anatomical context and confidence scores (F-beta, precision, and recall) calculated using NS-Forest (Liu et al. 2024), providing an objective measure of how useful marker sets are in identifying a cell type within the reference dataset. By selecting sets of markers, NS-Forest can provide the highest possible confidence marker sets for a given tissue context. Given that NS-Forest marker sets vary considerably in their F-beta scores, depending on data quality, cluster separability, and biological context, we provide the F-beta scores alongside the marker sets so that users can make informed decisions in their own analytical context. Wherever possible, we also fold transcriptomic classification into CL using the schema established for PCL (Tan et al. 2023)). Figure 6 shows an example and also shows how single cell transcriptomics can inform classification. Two transcriptomically distinct populations of GABAergic cortical interneurons express LAMP5. Classifying based on properties, we could define a Lamp5 neuron as any CGE derived cortical GABAergic interneuron that expresses LAMP5. However, doing so would define a class that subsumes Lamp5 Lhx6 neurons. In this case and for transcriptomic types in general, the transcriptomic hierarchy is primary.

Naming of T-types is often challenging. Where reference datasets are the result of dataset integration efforts, as is the case for the Human Cell Atlas integration efforts including the Human Lung Cell Atlas (Sikkema et al. 2023)), or naming is based on extensive annotation transfer, as in the Brain Initiative, the variety of names used in different source datasets need to be reconciled. In some cases, a large number of cell subtypes are identified, as is the case for amacrine cell subtypes in the Human Retina Atlas (Li et al. 2023), where CL editors are using cell type-specific marker genes from NS-Forest (Liu et al. 2024) combined with the parent cell type in the ontology as an objective means for cell type naming. CL editors are actively involved in many of these naming efforts as part of annotation jamborees. A successful example of this approach is the Breast Atlas Jamboree organized by CZI, which facilitated data integration across studies (Reed et al., 2024). This effort resulted in clusters that were previously named differently being unified under a single label approved by the mammary gland community (Gray et al. 2025). Annotation of single reference datasets rarely provides names that are clear and unambiguous in the broader context of cell types defined in CL. In these cases, CL keeps names of cell types from reference datasets as synonyms and official symbols and generates longer, more informative official names that frequently include transcriptional biomarkers for these T-types.

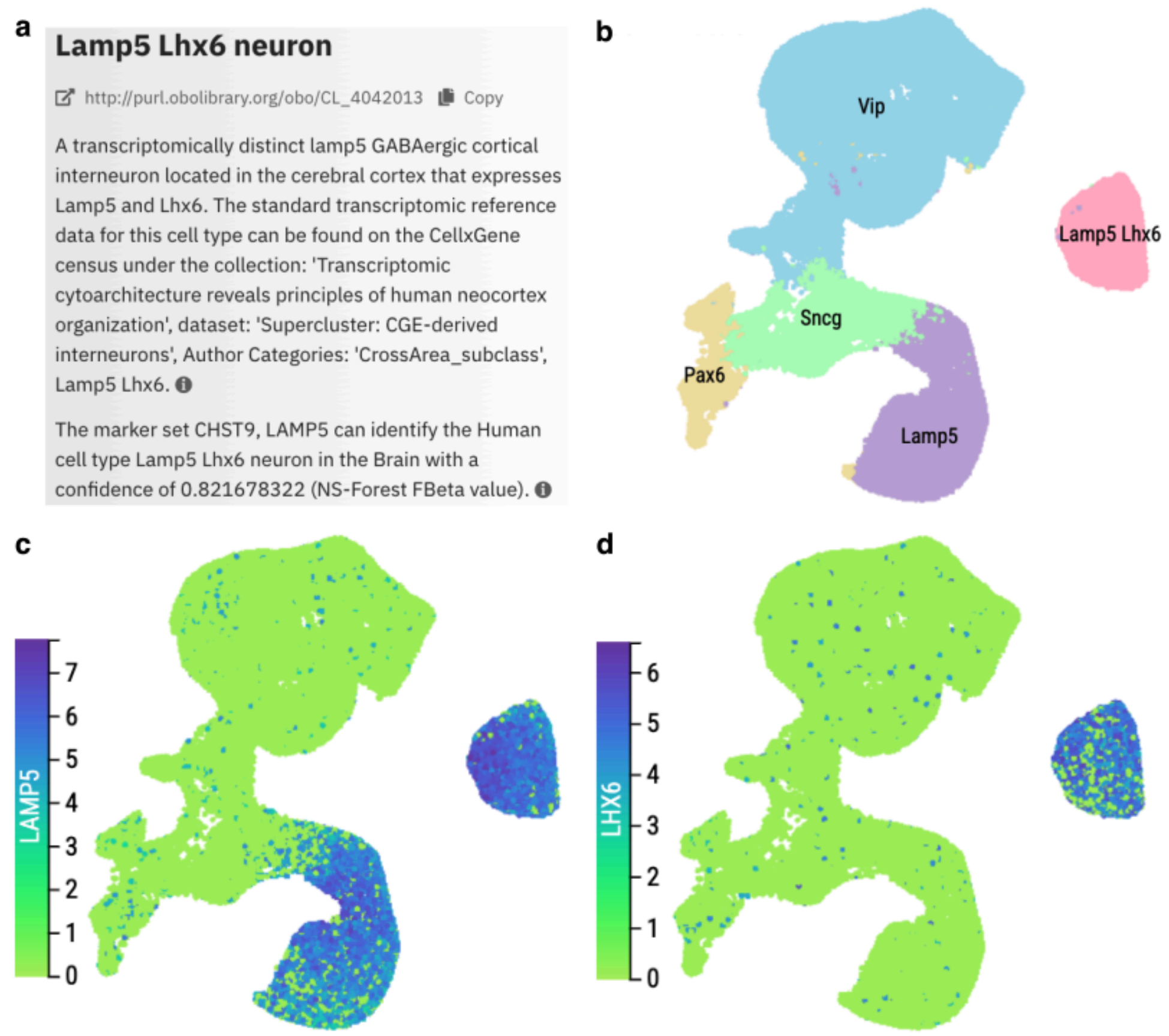

**Figure 6. Panel A** shows the definition of a T-type (cell type defined in part by a transcriptomic reference) Lamp5 Lhx6 neuron. Note the actionable link to transcriptomic data (linkouts include a dataset version specific link to data on CELLxGENE, not shown) and the NS-Forest markers derived from data (Peng 2025), confidence score and relevant tissue context. **Panel B:** UMAP of transcriptomic clusters in the reference dataset: CGE derived cortical interneurons from integrated samples from across the human cortex (Jorstad et al. 2023). **Panel C:** LAMP5 expression. **Panel D** LHX6 expression. Note that Lamp5 Lhx6 forms a distinct transcriptomic cluster that is not a subset of LAMP5 expressing cells.

## Taxon constraints

As CL aims to cover cell types across broad ranges of species, care must be taken to ensure that the ontology does not assert anything about a cell type that is not valid in all the species where that cell type is known to exist.

For example, ‘taste receptor cell’ (CL:0000209), which is defined as a chemoreceptor cell involved in the sensory perception of taste, is a very generic cell type that has a broad applicability across the animal kingdom (many animal species have such cells). It would therefore be wrong to assert (as has been the case in earlier versions of CL) that taste receptor cells are part of the ‘tongue’ (UBERON:0001723), because not all species that have taste receptor cells have them in the tongue, or even have a tongue to begin with. Notably, arthropods do have taste receptor cells, yet lack a tongue, which is an anatomical feature specific to chordates.

To facilitate the detection of such incorrect assertions, CL formally describes whenever possible the known taxonomic constraints (Deegan née Clark et al. 2010) that apply to a cell type, such as “this cell type is known to exist in that taxon”, “this cell type is known *not* to exist in that taxon”, or “this cell type is known to be *specific* to that taxon” (see Methods for details on how those constraints are formalized). This makes it possible for a reasoner to flag any assertion that would violate said taxonomic constraints, and thereby allows CL editors to fix such an incorrect assertion before it is even committed to the ontology.

### CL bridges classical definitions and transcriptomic definitions

Links between GO and CL are an important part of defining terms in both ontologies. In CL they constitute a formalised part of the classical definition, defining key functions and cell components. By combining annotation of genes to these GO terms with transcriptomic data, we can extract the transcriptomic signatures of these functions and components. We ran this analysis across markers from the majority of which come from single cell data. We assessed markers from CELLxGENE, based on automatically generated transcriptomic data annotated with CL terms (CZI Single-Cell Biology Program et al. 2023) and CellMarker2.0 (C. Hu et al. 2023), supported by high confidence annotation of genes to GO terms (those supported by experimental or evolutionary evidence, see methods for details) with direct links to CL terms. This approach finds transcriptomic signatures (600 markers in total; 315 from CELLxGENE) for 252 distinct structures and functions of 136 cell types supported by 428 references. An example of this is shown in Figure 1. Complete results are shown in supplementary tables S1 and S2 (https://drive.google.com/drive/folders/1PaDYkEtXH8nh-Q3Am129Am7Wm4NIaHnp).

In future we will extend this analysis to broader gene expression from transcriptomics data, concentrating on well-validated reference datasets to provide more general transcriptomic signatures of defining structures and functions of cell types. We will also extend it to take advantage of the broader semantics of GO, for example leveraging relationships between Cellular Components and Biological Processes for their assembly and maintenance and functions. We will also make it a priority to extend GO-CL links, leveraging agentic AI approaches to make this efficient. This will make an increasingly rich set of transcriptomic profiles for the functions of cell type defined in CL available through the CL knowledge graph or via independent analysis.

# Discussion

### Data-linked definitions

As cell typing becomes more data driven, CL faces new challenges in defining and naming cell types in a way that is relevant to data-driven annotation. CL and its PCL extension have risen to this challenge by developing and using formal data-linked cell-type definitions: computationally tractable links to reference data that can be used to retrieve and analyze this data and use it as a reference in automated annotation. At a minimum this requires resolvable links between ontology terms and files containing the relevant reference data and mechanisms to refer to specific defining data for a cell type. Data archives such as Nemo (Ament et al. 2023) currently lack sufficient granularity for referencing files, and standardization of files is not always sufficient to unambiguously reference data within them. The development of resources, such as CELLxGENE Census, that harmonize representation of single cell transcriptomics data across large numbers of datasets and provide APIs for accessing data provides easy solutions to this problem, as long as the reference data is deposited on such a resource.

Choosing reference data is also a challenge. Reference standards need to be sufficiently stable—a reference that changes rapidly in radical ways without clean mappings and migration paths between versions is not suitable for supporting reliable searching and browsing in integrative resources such as CELLxGENE Discover. For this reason, we have mostly so far restricted these types of definitions to the PCL. However, standard and relatively stable reference datasets are beginning to emerge from the work of the Human Cell Atlas integration efforts and the BRAIN Initiative. For example, CL uses standard references for gross cell types in the human cortex (Jorstad et al. 2023) to define cortical cell type via a combination of textual definitions and formal links to data hosted on CELLxGENE Discover (see Figure 6).

## T-type hierarchy and naming

Decisions on nomenclature are challenging for any ontology, but the rise of single-cell transcriptomics, the use of clustering in latent space to define T-types, and the increasing use of annotation transfer have brought new challenges. Classically defined cell types are typically named for their intrinsic and extrinsic properties. For example, L4 IT neurons in the cortex were originally named for their intra-telencephalic projection pattern (IT) and soma location in cortical layer 4 (L4). Similarly, VIP interneurons in the cortex were originally named for their expression of Vasoactive Peptide. However, recent papers have published T-types with L4 IT in the name that refer to neurons in regions with no layer 4 (Jorstad et al. 2023; Bakken et al. 2021), and T-types with VIP in the name where VIP expression is no higher than in many interneuron types not named VIP (Yao et al. 2020) . These names make perfect sense based on annotation transfer and clustering, but potentially violate community expectations around the meaning of names.

This poses a challenge for standard ontology definitions, where properties in the name are typically reflected in properties asserted for members of the class. One potential solution, trialed in PCL, is to append '-like' to terms defined by links to reference transcriptomics that have subclasses that do not share the properties in the name. For example we could use L4 IT-like for the L4-IT T-type named above. However, this results in terms that do not reflect usage. Alternatively, CL can accept that names do not necessarily reflect actual properties and expression, acknowledging this in the formal structure of the ontology (only linking VIP expression to T-types where expression is above background) and making exceptions clear in comments attached to terms.

Transcriptomic types are also frequently organised into hierarchies based on transcriptomic similarity. This might be overall transcriptomic similarity based on highly variable genes or a more restricted subset of gene expression, such as transcription factor expression profiles (Bakken et al. 2021; Yao et al. 2023). These hierarchies are arguably more scientifically grounded than traditional property based classification and so it is important to represent them in cell ontologies. PCL already represents many such hierarchies representing assertions of homology between cell types using simple pairwise relationships (Tan et al. 2023). With the emergence of increasing consensus on types in the mammalian brain, CL plans to integrate standard reference T-types and hierarchies from BICAN.

However, there are dangers in assuming that these hierarchies reflect simple subClassOf hierarchies down which properties are inherited. Markers that can be used to identify a very general T-type with high confidence may not be expressed in rare types nested under them in the hierarchy. Similarly, assertions about location or function based on mapping to spatial transcriptomics and patch-seq to one level in a taxonomy may not hold for all subtypes. This can be solved by moving many of these assertions to non-logical relationships (annotation properties), which are not inherited down the subclass hierarchy, or by replacing subClassOf relationships between hierarchical T-types with a non-logical relation that specifically reflects transcriptomic hierarchy.

## Bridging modalities with data

Our understanding of the degree of correspondence between transcriptomic and classically defined cell types is likely to increase with the use of multimodal techniques that provide a link to anatomy, morphology, and function. For example, location plays a critical role in defining and mapping many cell types in a combined single cell and spatial transcriptomics atlas of the mouse brain—classical markers that are insufficient to identify cell types in the context of the whole brain effectively identify them in a specific anatomical location (Figure 2) (Scala et al. 2020; Yao et al. 2023). Similarly, morphology plays a critical role in defining and mapping cell types in a combined patch-seq and single cell atlas of the mouse motor cortex (Cadwell et al. 2016; Scala et al. 2020).

## Challenges of harmonizing cross-species cell types in CL

General cell types are typically very well conserved across vertebrates even when their subtypes show species-specific specialization. This is most apparent in organs whose function is under strong selective pressure. For example, the retina is specialised in different mammals to support the acuity, color vision, motion tracking, contrast detection and energy efficiency required for their survival (Collin and Trezise 2004; Shekhar and Sanes 2021; Hahn et al. 2023). The three-layer structure of the retina and its major cell types (rod and cone photoreceptors, Müller glia, amacrine cells, bipolar cells, retinal ganglion cells and horizontal cells) is conserved across vertebrates (Hahn et al. 2023). However, there is significant variation in the structure, function, distribution, and connectivity of subtypes of these general classes particularly among the amacrine and retinal ganglion cell (RGC) subtypes.

Until recently, we had limited ability to map these more specialised cell types across species, relying on morphology, function and a limited set of markers to do so (Masland 2012; Zeng and Sanes 2017). As a result, there is a proliferation of RGC terms in CL derived from studies of single species, with only very limited cross species grouping. There are 68 RGC terms in total, the majority from studies in mice, but with non-overlapping sets of terms for rabbits and primates (including humans).

Single cell transcriptomics now provides data-rich evidence for cell type homology across species. For example, Hahn et al. (Hahn et al. 2023) define RGC "orthotypes" based on single cell RNAseq analysis of 12 different mammalian species. Using this approach they find strong evidence that the midget ON & OFF RGCs of primates are homologous to much rarer cell types found in the mouse retina (c42 and c43) that have decreased in size and increased in number as primates evolved. Similarly, the Brain Initiative Cell Atlas Network is generating cross-species consensus atlases integrating human and non-human primate brains, mapping types to mice where possible. This has resulted in the recent publication of a cross-species consensus atlas of neuron types in the Basal Ganglion (Johansen et al. 2025).

CL has recently added cross-species types, defined in part via links to reference atlas for both RGCs and basal ganglion neurons using standard formal schemas for linking (Tan et al. 2023). However, we exercise caution in interpretation: while transcriptomic similarity provides strong evidence for homology, it could reflect convergent evolution, particularly between distantly related taxa. These grouping terms link to the data and analyses supporting homology assertions, representing an initial step toward distinguishing homologous and analogous cell types without yet encoding this distinction explicitly in the ontology structure.

## Markers

While CL has always supported recording cell type markers, the proliferation of markers from omics data and their dependence on tissue context assay and analysis algorithms means that it is increasingly challenging to record markers consistently and comprehensively enough within CL to meet community needs without bloating CL content.

LLM pipelines show some promise in direct annotation of cell types with reasonable free text terms (Hou and Ji 2024). However, these lack the advantage of controlled annotation, are currently only accurate for very general cell types and are potentially prone to hallucination in the absence of curation.

To support community need for queries of markers linked to CL, we are building a Knowledge Graph extension to the Cell Ontology that folds in multiple sources of markers, including our own LLM aided curation, as well as anatomical and experimental context, evidence and provenance, including links to reference data and data driven validation via links to analysis of data on CELLxGENE. Where available, we will also fold in NS-Forest markers (Aevermann et al. 2021; Liu et al. 2024), which provide machine learning-based optimal combinations for identification of markers in the context of a dataset, along with a confidence score for marker combinations relative to the dataset that they are derived from. This, we believe, provides a practical and objective standard for markers, while allowing access to a broad range of markers used by the community.

The knowledge graph API will support searching for cell type markers by anatomical context, taking advantage of Uberon structure to support a wide range of granularities. Results will be ranked based on the weight of supporting evidence and can be filtered for official sources (e.g. HCA official integration markers). The provenance model employed will support full attribution and links to all original sources.

## Integration of large language models

Large Language Models (LLMs) are now a core component of our ontology editing pipelines. Editors primarily use Deep Search (Xu and Peng 2025) to generate draft cell type definitions with literature references, then review the output—checking the references, removing unsupported or irrelevant assertions and ensuring definitions align with CL standards. Rather than relying on one reference at a time, LLMs can quickly summarise current knowledge across multiple sources around a cell type, speeding up the drafting process. With well-engineered prompts, editors can produce definitions that closely follow the standards of CL. However, research on LLM-generated definitions shows that users with lower domain expertise are more likely to accept a generated definition on face value, even if incorrect (Toro et al. 2024), so expert review remains essential. While Deep Search minimises LLM hallucinations, this can be further reduced by integrating it with Retrieval Augmented Generation (RAG), which automatically checks generated referenced text for consistency with references (Skarlinski et al. 2024).

To directly address this, we have developed **CLARA (CL Accuracy Reference Agent) (https://github.com/Cellular-Semantics/clara-agent)**, an agentic pipeline built using PaperQA2 (Skarlinski et al. 2024) that validates both formal and textual definitions against their supporting references. CLARA automatically retrieves references associated with a candidate term, breaks the text into individual assertions, and checks whether each assertion is supported or contradicted by the literature. In an initial pilot evaluation of 689 assertions, CLARA showed high precision and recall (over 0.9 for both). Future work will focus on more extensive validation of CLARA and integrating CLARA into GitHub Actions so that every proposed new term or edit in CL—including those generated by LLMs—is automatically validated, with any errors flagged before an editor reviews. This provides an additional layer of quality control and significantly reduces the risk of hallucinated or unsupported content entering the ontology. Importantly, all content generated using LLM and agentic tools undergoes the same rigorous expert review process as the manually curated content prior to inclusion in CL.

Agentic AI is now fully integrated into our ontology editing workflows using GitHub Copilot (Figure 7). These agentic tools operate within GitHub to interpret issues created by editors and automatically generate OWL edits via Pull Requests (PRs), allowing editors to focus on review rather than drafting changes from scratch. To date, 73 PRs created by GitHub Copilot have been merged into CL, showing how agentic tools can speed up routine ontology editing. Although they are currently used for well-defined tasks, these tools have the potential to handle broader, higher-level instructions and turn them into accurate, context-appropriate ontology updates as the technology continues to mature. In parallel, we are exploring multi-agent frameworks such as Aurelian (https://github.com/monarch-initiative/aurelian), which coordinate complex tasks such as literature review, semantic annotation, and ontology navigation. These emerging approaches may help automate literature extraction for new cell types, propose candidate definitions and mappings, and support consistency checking across ontologies such as GO and Uberon. Ultimately, adopting scalable, agent-assisted curation workflows will be key for future data integration, allowing CL to evolve quickly and keep pace while providing accurate, standardised grounding for the rapidly growing number of cell types emerging from single-cell atlases.

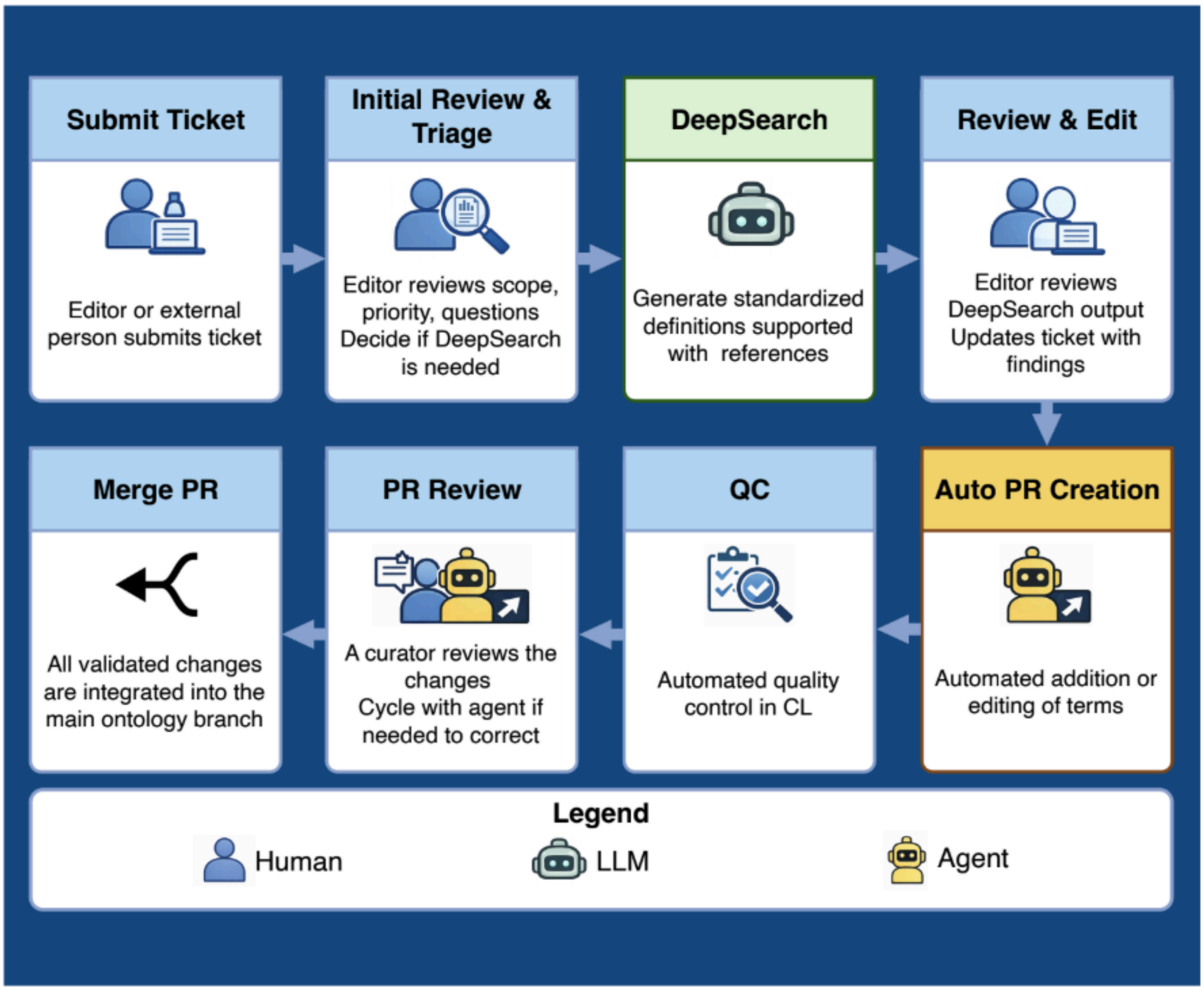

**Figure 7.** Workflow for the creation of new ontology terms. A ticket is submitted by an editor or external contributor describing the proposed term(s). An editor then performs an initial review and triage to assess scope, priority, and outstanding questions, and to determine whether automated literature support is required. When appropriate, Deep Search is used to generate standardized draft definitions supported by literature references, which are subsequently reviewed and edited by the editor. Once all required details are present in the ticket, an AI agent implements the changes in the ontology and creates a Pull Request (PR). The ontology edits then undergo automated quality control to ensure that no structural or semantic errors are introduced that could affect reasoning or downstream data integration. Finally, a human curator reviews the PR and iterates with the agent until the changes are approved and merged into the main ontology branch.

## Future prospects

CL is a critical resource for making data about cell types FAIR. It provides consistent, well-defined cell type terms for large scale atlasing efforts, in particular those built on transcriptomics. By providing stable identifiers with machine readable axiomatization, CL provides the underpinnings for semantic data products that can be reliably reused, combined, and queried. Such semantically aligned data products not only enable symbolic reasoning, but are particularly well suited for connectionist AI systems which benefit from input data being structured, aligned, and grounded in ontologies (Sahoo et al. 2022; Kulmanov et al. 2021). Bridging classical and transcriptomic definitions it distills key prior knowledge about cell types using both well-refenced text and formal assertions (e.g. via links to the Gene Ontology. Increasing adoption of agentic workflows to edit CL combined with extensive community engagement are rapidly improving the scalability of CL development and its coverage of biology.

Rapid advances in Machine Learning (ML) and agentic AI are transforming the landscape of single cell research, bringing a constantly evolving set of new tools supporting data integration, analysis and interpretation. This generates new challenges and opportunities for CL use and further development.

Major training datasets for Machine Learning are ready-annotated with CL terms, the most prominent example being the CELLxGENE corpus. AlphaGenome (Avsec et al. 2026) is a model for predicting cell type and tissue specific prediction of regulatory effects on mutations, and uses the CL both in the preparation of training data, and at inference time. New ML models built on multi-modal embeddings identify cell types, the tissue 'niches' in which they are located and both known and uncharacterized gene programs that underpin their structure and function (Acosta et al. 2025; Zhao et al. 2025; Schaar et al. 2024; Birk et al. 2025). At least some of these models use morphology and transcriptomics to define cell types (Acosta et al. 2025) arguably reflecting a more classical approach to cell type definition than recent work in single cell transcriptomics.

New agentic frameworks are capable of running analysis and annotation pipelines and aiding atlas interpretation using information mined from the literature and public databases and using gene enrichment analysis to map to pathways and biological processes (Ahuja et al. 2025; Wang et al. 2025; Nouri et al. 2025). At least some also use the cell ontology directly for annotation (Wang et al. 2025; Ahuja et al. 2025) or indirectly via tools that do (Nouri et al. 2025).

With all these advances, the need for high-quality standard frameworks as a locus of integration for data and prior knowledge remains. CL is uniquely well positioned to play this role while the workflows developed to extend and adapt it will ensure it has sufficient coverage and quality to remain relevant.

# Methods

## Implementing human readable diffs on GitHub PRs

In order to simplify the review process, a GitHub actions automation automatically posts a comment on pull requests into the main branch, when given a specific command in the comment, that shows a human-readable report of differences between both unreasoned and reasoned files in the pulling branch and the main branch, prepared using a WHELK reasoner (Balhoff and Mungall 2024)). This is achieved by using the ROBOT “diff” function through the ODK on GitHub actions (Jackson et al. 2019; Matentzoglu, Goutte-Gattat, et al. 2022).

## Implementing taxon constraints

CL, in conjunction with the Uberon anatomy ontology, makes use of a handful of properties to formally represent known constraints between a cell type and a taxon. They are (1) “in_taxon” (RO:0002162), used to indicate that a cell type is specific to a given taxon; (2) “never_in_taxon” (RO:0002161), used to indicate, conversely, that a cell type does not exist in a given taxon; and (3) “present_in_taxon” (RO:0002175), used to indicate that a cell type is known to exist in a given taxon, but without implying, contrary to RO:0002162, that it is specific to that taxon.

Those properties are intended to be used conjointly with a specific variant of the NCBI Taxonomy ontology that contains additional disjointness axioms between sibling taxa, and that can be used by an EL reasoner to detect whenever taxon constraints are violated. RO:0002161 and RO:0002175 are annotation properties that are “expanded” at build-time into logical axioms by ROBOT’s “expand” command, according to a principle similar to OWLTools’ macros (Mungall et al. 2011). An annotation assertion like “CL:0002174 never_in_taxon NCBITaxon:50577” (“follicular cells of ovary don’t exist in insects”) is expanded into the following two axioms:

CL:0002174 DisjointWith: in_taxon some NCBITaxon:50577
CL:0002174 SubClassOf: in_taxon some (not NCBITaxon:50577)

Conversely, an annotation assertion like “CL:0000711 present_in_taxon NCBITaxon:9606” (“cumulus cells are known to exist in humans”) is expanded into a “witness class” *W* such as

*W* SubClassOf: CL:0000711
*W* SubClassOf: in_taxon some NCBITaxon:9606

The purpose of that witness class is solely to deliberately cause an incoherency in the ontology, should the taxon constraint ever be violated (for example, if CL:0000711 ended up being mistakenly classified as a subclass of a cell type that is specific to a taxon disjoint with humans).

## Integration with taxon-specific ontologies

CL maintains mappings between its own terms and equivalent terms in taxon-specific ontologies. Historically and in many OBO ontologies, inter-ontology mappings have been represented by *oboInOwl:hasDbXref* annotations (aka “cross-references”). Recently, the Simple Standard for Sharing Ontological Mappings (SSSOM) has been devised to represent cross-ontology mappings without the shortcomings of cross-references (Matentzoglu, Balhoff, et al. 2022). The *Drosophila* Anatomy Ontology (DAO) has already switched to using SSSOM for its mappings with Uberon and CL, and CL has been updated to automatically fetch and use the DAO-provided, SSSOM-formatted mapping set. SSSOM brings several benefits over the old cross-reference format, such as the possibility to use meaningful mapping predicates from the SEMAPV vocabulary (Matentzoglu, Flack, et al. 2022), specifically intended to represent cross-species mappings, and the possibility to annotate each mapping with rich provenance metadata. We are currently using the DAO-CL mappings as a testbed for using SSSOM-maintained mappings, and we hope to progressively generalize the use of SSSOM to the mappings between CL and all the other taxon-specific ontologies.

Regardless of their formats (cross-references or SSSOM), all mappings are ultimately used to generate “bridging axioms” between CL and the taxon-specific ontologies. For example, a mapping between CL term for ‘neuron’ (CL:0000540) and the corresponding term in DAO (FBbt:00005106) is derived into an equivalence axiom of the form *FBbt:00005106 EquivalentTo: CL:0000540 and (part_of some NCBITaxon:7227)*, in effect stating that a fly neuron (FBbt:00005106) is a generic neuron (CL:0000540) that is part of a fruit fly (NCBITaxon:7227). This ensures that when CL and DAO are merged together, the DAO term is correctly positioned within the resulting merged hierarchy (Figure 8). We expose these mappings in a number of ways in order to provide infrastructure for user facing applications. Firstly, we release a combined product “Composite Metazoan”, a single unified anatomy and cell type ontology that integrates a dozen of taxon-specific ontologies into the taxon-neutral framework of CL and Uberon. As such, Composite Metazoan is a convenient, readily available option for applications that require broad coverage of the animal kingdom. Second, we release taxon-specific bridge files, each containing the bridging axioms for a given taxon — these may be used by a third-party application to build an integrated cross-species ontology covering precisely the taxa that the application required. Lastly, we also publish the mapping themselves, as a single SSSOM (Matentzoglu, Balhoff, et al. 2022) mapping set, for downstream applications that require more flexibility than what the pre-made bridge files can provide.

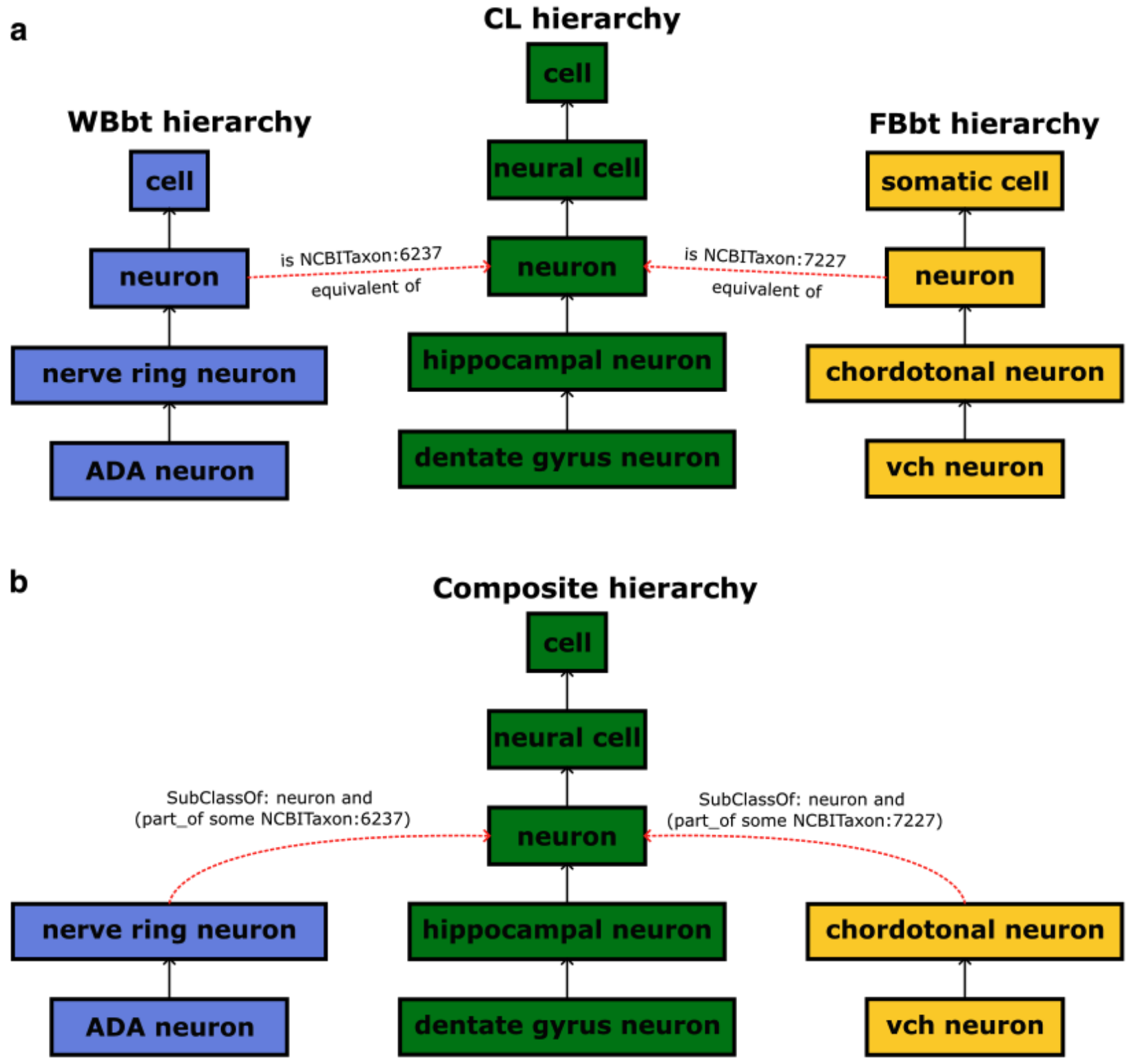

**Figure 8.** CL and the species-specific anatomy ontologies each maintain their own hierarchy of terms separately. (Top) The hierarchies are bridged by means of cross-species equivalence axioms that connect species-specific terms to their CL species-neutral counterparts. (Bottom) A single, unified hierarchy is obtained by replacing the species-specific terms by class expressions that use the appropriate species-neutral term from CL.

# Generating CL textual definitions using LLMs

The improvements in the tuft cell hierarchy illustrated above was supported by LLMs. A standardised prompt was created, tested and refined to align with CL standards. This prompt is input into the default Perplexity Pro model to generate textual definitions for a given {cell type}(https://www.perplexity.ai/search/system-role-you-are-an-expert-V1D0txTvQxuE0ApqaG3ZTw). Editors may run the prompt multiple times for a better output. Editors then assess the LLM-generated definitions and supporting references, refining the definition accordingly. During manual editing, editors focus on removing redundant general information, validating assertions against provided references, and including species-specific information (with references) if unique markers are specified. An example of thymic tuft cell definition is shown above (Figure 5). The text below represents the standardised prompt currently in use by CL to create LLM-generated textual definitions.

**System Role: You are an expert cell biologist with extensive experience in creating precise and informative descriptions of cell types for ontologies.**
**User Role: I need you to create definitions for specific cell types to be included in the Cell Ontology.**
Each definition should:
1. Avoid naming the cell type being defined directly. It should start with a statement of a general classification for the cell type being defined, followed by the characteristics that distinguish it from other cell types within the same general classification.
2. Describe distinguishing characteristics, including structural features, functional roles, and anatomical context.
3. Include species-specific information when relevant, noting presence or absence in different organisms.

4. Mention key molecular markers, transcription factors, or genes only if they are crucial for identification or development of the cell type. When including molecular markers, specify the species in which they have been identified (e.g., "marker X in mice", "marker Y in humans").
5. For general cell types, focus on common features across different tissues or organs.
6. Include supporting references to key statements of the definition rather than listing them at the end.
7. Be concise yet comprehensive, aiming for 80-120 words in a single paragraph.
8. Use clear, scientific language accessible to biologists across various specialties.

Example output for a specific cell type: "A tuft cell that is part of the epithelium of pancreatic duct. Present in humans and rats, this cell is absent in the murine pancreas under normal conditions but emerges during acinar-to-ductal metaplasia triggered by injury, inflammation, or oncogenic mutations (1,2). It modulates the immune response and protects against pancreatic ductal adenocarcinoma progression by producing suppressive eicosanoids, such as prostaglandin D2 (3). A tuft cell in the pancreatic duct highly expresses the transcription factor POU2F3, which is essential for its development and presence (4)."
**"Please generate a definition for the following cell type: 'tuft cell of thymus'"**

## CL-Knowledge graph and GO-Marker cross-analysis

The Ontology Based Application Starter Kit, OBASK (Kir et al. 2024), is a well-tested, containerised pipeline that supports conversion of OWL ontologies and RDF into annotated nodes and edges in a Neo4J graph using a standard set of mappings from OWL axioms that relate only 2 entities (classes, individuals). We used this pipeline to generate a graph representing a merge between the Cell Ontology (including markers recorded in CL) and the Gene Ontology, including relationships between terms in these ontologies in both directions.

We then folded additional edges into this graph representing links between cell types and their markers and between GO terms and genes annotated to them, focusing on GO terms that are directly related to CL terms in the graph.

More specifically, we added gene:cell type edges representing markers from CELLxGENE (generated from the CxG corpus of transcriptomic data annotated with CL (CZI Single-Cell Biology Program et al. 2023)), CellMarker2.0 (C. Hu et al. 2023) and the Scheurmann lab (Hu et al. 2025). We queried the QuickGO API (Munoz-Torres and Carbon 2017) to identify all mouse and human genes annotated to the selected GO terms, where GO annotation used the evidence codes 'biological aspect of ancestor evidence used in manual assertion' (ECO:0000318) and 'experimental evidence used in manual assertion' (ECO:0000269) or their descendant codes. These were loaded into the graph as GO-Gene edges.

Finally, we used Cypher to query the resulting knowledge graph for genes that are both markers of a specific cell type and implicated in a GO process or Cell Component directly linked to that cell type.

Keeping this updated with source ontologies and data sources simply requires re-running the pipeline.

# Data Availability

CL can be accessed via a variety of routes. It can be browsed on the Ontology Lookup Service ((Jupp et al. 2015; McLaughlin et al. 2025) and Bioportal (Vendetti et al. 2025; "Cell Ontology in Bioportal," n.d.). A stripped down version of CL with links to markers can also be browsed on CZ CELLxGENE CellGuide (see Figure 3).

CL is available for download from a set of standard persistent URLs in a variety of forms (see below) and formats (OWL, OBO and OBOgraphs JSON). The main CL release is available from:

http://purl.obolibrary.org/obo/cl.{obo/owl/json}
All other releases are available via
http://purl.obolibrary.org/obo/cl/cl-{form}.{obo/owl/json}

Each download also includes a version URL—a permanently resolvable link to the version that can be used to reference versions used in analysis and annotation.

e.g.
http://purl.obolibrary.org/obo/cl/releases/2025-01-08/cl.owl

Downloads available by content (form):

- cl - The Cell Ontology plus merged imports, including from Uberon and the Gene Ontology
- cl-plus - combines the full Cell Ontology with the provisional cell ontology
- cl-base - The Cell Ontology with links to external ontologies recorded as bare IRIs. This is designed to be combined with other base ontologies to make integrated products and for use in generating import modules from CL.
- cl-simple - only CL terms and their relationships. No imports
- cl-basic - only CL terms and their relationships. Additionally the graph is guaranteed to lack cycles (these are perfectly legal in OWL but removing them makes it easier for some software to operate on it).

Each release also comes with a detailed, automated report of changes since the last release.

CL is also available as part of 'composite-metazoan'—an ontology that combines, Uberon, CL and species-specific ontologies: Drosophila Anatomy Ontology (FBbt); *C. elegans* Gross Anatomy Ontology (WBbt); Zebrafish Anatomy and Development Ontology (ZFA); Xenopus Anatomy Ontology (XAO); Mouse Adult Gross Anatomy (MA) and Mouse Developmental Anatomy Ontology (EMAPA); Human Developmental Anatomy (EHDAA2); and the ontologies derived from the Allen Institute's brain atlases. The composite ontology is available at http://purl.obolibrary.org/obo/uberon/composite-metazoan.owl.

Programmatic access is available via the Ontology Lookup Service API (https://www.ebi.ac.uk/ols4/api/ontologies/cl/), the Bioportal API (see http://data.bioontology.org/documentation), via the UberGraph (Balhoff et al. 2022) SPARQL endpoint (https://ubergraph.apps.renci.org/sparql), and via the Ontology Access Kit (OAK). UberGraph supports queries of all asserted and inferred relationships between terms in CL and terms in a broad range of OBO-Foundry ontologies that reference CL, including the Gene Ontology, Uberon, Human Phenotype Ontology and MONDO. A REST API is also available, featuring pre-rolled template queries of Ubergraph, for example allowing users to query for all cell types in a specific anatomical structure (https://grlc.io/api-git/INCATools/ubergraph/subdir/sparql).

MIRACL sheet that uses a search API can be found in here https://zenodo.org/records/15387766 (Bayindir 2025)

# Code Availability

The Cell Ontology is developed, documented and released via a GitHub repository (https://github.com/obophenotype/cell-ontology) implemented using the Ontology Development Kit (Matentzoglu, Goutte-Gattat, et al. 2022). This includes all code and ontology files used in the development of the ontology, available under an open license, as well as a tracker for community requests. Code and data used for stats in this paper can be found at https://github.com/shawntanzk/cl-manuscript.

All code used to build the knowledge-base can be found at https://github.com/Cellular-Semantics/CL_KG.

# Author Contributions

SZKT, APB, DGG, CE, ADD, and DOS wrote the manuscript, and all authors reviewed the manuscript. All authors contributed to the development of CL and/or major tools that use CL and are described in the manuscript

# Funding

APB and ARC were funded by NIH via the HuBMAP project (OT2OD033756 and OT2OD026671) and EMBL-EBI core funds. CE and DOS contribution was supported by grant number 2023-221471 from the Chan Zuckerberg Initiative DAF. JPB was partly supported by the GO consortium that is funded by the National Human Genome Research Institute (NHGRI), US National Institutes of Health, grant number HG012212. IUB and JJB's work was funded by a gift from Schmidt Futures (https://www.schmidtfutures.org) to support the 'Cell Annotation Platform'. MAH, ST, and RS were supported by NIH Office of the Director Grant #5R24OD011883 for the Monarch Initiative. SZKT, HK, ADR & DOS work was supported in part by NIH 1UM1MH130981-0. TL was supported by a grant from the São Paulo Research Foundation (#19/26284–1). ADD, JAO, and BP were supported in part by the National Institute of Allergy and Infectious Disease [4U19AI118610]. CJM was supported by the National Institutes of Health National Human Genome Research Institute [HG010860, HG012212, HG010859]; National Institutes of Health Office of the Director [R24 OD011883]; and the Director, Office of Science, Office of Basic Energy Sciences, of the US Department of Energy. PR was supported by a grant from Chan Zuckerberg Initiative DAF to support the Human Cell Atlas data coordination platform. YZ was supported, in part, by the Division of Intramural Research of the National Library of Medicine (NLM), National Institutes of Health. RS was supported by the NIH National Human Genome Research Institute Phenomics First Resource, NIH-NHGRI # 5RM1 HG010860, a Center of Excellence in Genomic Science. The work of DOS, HK, UB, JB & CE was funded in part by 220540/Z.20/A, “Wellcome Sanger Institute Quinquennial Review 2021-2026. DGG was supported by grant BB/T014008 from the UK Biotechnology and Biological Sciences Research Council (BBSRC) and the US National Science Foundation Directorate of Biological Sciences (NSF/BIO).

# Competing Interests

MAH is the founder of Alamya Health.

# Acknowledgements

We thank Aidan Maartens (Wellcome Sanger Institute) for scientific writing support.

# References

Acosta, Paul H., Pingjun Chen, Simon P. Castillo, Maria Esther Salvatierra, Yinyin Yuan, and Xiaoxi Pan. 2025. "CellSymphony: Deciphering the Molecular and Phenotypic Orchestration of Cells with Single-Cell Pathomics." In *arXiv [cs.CV]*. August 13. arXiv. http://arxiv.org/abs/2508.10232.

Aevermann, Brian, Yun Zhang, Mark Novotny, et al. 2021. "A Machine Learning Method for the Discovery of Minimum Marker Gene Combinations for Cell Type Identification from Single-Cell RNA Sequencing." *Genome Research* 31 (10): 1767–1780.

Ahuja, Gautam, Alex Antill, Yi Su, et al. 2025. "Multi-Agent AI Enables Evidence-Based Cell Annotation in Single-Cell Transcriptomics." In *bioRxiv*. November 7. https://doi.org/10.1101/2025.11.06.686964.

Ament, Seth A., Ricky S. Adkins, Robert Carter, et al. 2023. "The Neuroscience Multi-Omic Archive: A BRAIN Initiative Resource for Single-Cell Transcriptomic and Epigenomic Data from the Mammalian Brain." *Nucleic Acids Research* 51 (D1): D1075–D1085.

Artis, David, and Hergen Spits. 2015. "The Biology of Innate Lymphoid Cells." *Nature* 517 (7534): 293–301.

Avsec, Žiga, Natasha Latysheva, Jun Cheng, et al. 2026. "Advancing Regulatory Variant Effect Prediction with AlphaGenome." *Nature* 649 (8099): 1206–1218.

Baek, Seungbyn, and Insuk Lee. 2020. "Single-Cell ATAC Sequencing Analysis: From Data Preprocessing to Hypothesis Generation." *Computational and Structural Biotechnology Journal* 18 (June): 1429–1439.

Bakken, Trygve, Lindsay Cowell, Brian D. Aevermann, et al. 2017. "Cell Type Discovery and Representation in the Era of High-Content Single Cell Phenotyping." *BMC Bioinformatics* 18 (Suppl 17): 559.

Bakken, Trygve E., Nikolas L. Jorstad, Qiwen Hu, et al. 2021. "Comparative Cellular Analysis of Motor Cortex in Human, Marmoset and Mouse." *Nature* 598 (7879): 111–119.

Balhoff, James P., Ugur Bayindir, Anita R. Caron, Nicolas Matentzoglu, David Osumi-Sutherland, and Christopher J. Mungall. 2022. "Ubergraph: Integrating OBO Ontologies into a Unified Semantic Graph." *CEUR Workshop Proceedings*.

Balhoff, James P., and Christopher J. Mungall. 2024. "Whelk: An OWL EL+RL Reasoner Enabling New Use Cases." Preprint, Schloss Dagstuhl - Leibniz-Zentrum für Informatik, December 18. https://doi.org/10.4230/TGDK.2.2.7.

Balhoff, Jim, and Colin k. Curtis. 2021. *INCATools/ubergraph: Release 2021-03-26*. https://doi.org/10.5281/zenodo.4641309.

Bard, Jonathan, Seung Y. Rhee, and Michael Ashburner. 2005. "An Ontology for Cell Types." *Genome Biology* 6 (2): R21.

Bastian, Frederic B., Julien Roux, Anne Niknejad, et al. 2021. "The Bgee Suite: Integrated Curated Expression Atlas and Comparative Transcriptomics in Animals." *Nucleic Acids Research* 49 (D1): D831–D847.

Bayindir, İsmail Uğur. 2025. "Ontology-Autosuggest Metadata Annotation Spreadsheet." Preprint, Zenodo. https://doi.org/10.5281/ZENODO.15387766.

Bennett, Hayley M., William Stephenson, Christopher M. Rose, and Spyros Darmanis. 2023. “Single-Cell Proteomics Enabled by next-Generation Sequencing or Mass Spectrometry.” *Nature Methods* 20 (3): 363–374.

Birk, Sebastian, Irene Bonafonte-Pardàs, Adib Miraki Feriz, et al. 2025. “Quantitative Characterization of Cell Niches in Spatially Resolved Omics Data.” *Nature Genetics* 57 (4): 897–909.

Börner, Katy, Philip D. Blood, Jonathan C. Silverstein, et al. 2025. “Human BioMolecular Atlas Program (HuBMAP): 3D Human Reference Atlas Construction and Usage.” *Nature Methods* 22 (4): 845–860.

Börner, Katy, Sarah A. Teichmann, Ellen M. Quardokus, et al. 2021. “Anatomical Structures, Cell Types and Biomarkers of the Human Reference Atlas.” *Nature Cell Biology* 23 (11): 1117–1128.

Brügger, Michael David, and Konrad Basler. 2023. “The Diverse Nature of Intestinal Fibroblasts in Development, Homeostasis, and Disease.” *Trends in Cell Biology* 33 (10): 834–849.

Bueckle, Andreas, Bruce W. Herr II, Josef Hardi, Ellen M. Quardokus, Mark A. Musen, and Katy Börner. 2024. “Construction, Deployment, and Usage of the Human Reference Atlas Knowledge Graph for Linked Open Data.” In *Bioinformatics*, No. Biorxiv;2024.12.22.630006v1. BioRxiv, December 23. https://www.biorxiv.org/content/10.1101/2024.12.22.630006v1.

Cadwell, Cathryn R., Athanasia Palasantza, Xiaolong Jiang, et al. 2016. “Electrophysiological, Transcriptomic and Morphologic Profiling of Single Neurons Using Patch-Seq.” *Nature Biotechnology* 34 (2): 199–203.

Cao, Zhi-Jie, Lin Wei, Shen Lu, De-Chang Yang, and Ge Gao. 2020. “Searching Large-Scale scRNA-Seq Databases via Unbiased Cell Embedding with Cell BLAST.” *Nature Communications* 11 (1): 3458.

Caron, Anita R., Aleix Puig-Barbe, Ellen M. Quardokus, et al. 2024. “A General Strategy for Generating Expert-Guided, Simplified Views of Ontologies.” In *bioRxiv*. December 18. https://doi.org/10.1101/2024.12.13.628309.

Caufield, J. Harry, Harshad Hegde, Vincent Emonet, et al. 2023. “Structured Prompt Interrogation and Recursive Extraction of Semantics (SPIRES): A Method for Populating Knowledge Bases Using Zero-Shot Learning.” In *arXiv [cs.AI]*. April 5. arXiv. http://arxiv.org/abs/2304.02711.

Causeret, Frédéric, Matthieu X. Moreau, Alessandra Pierani, and Oriane Blanquie. 2021. “The Multiple Facets of Cajal-Retzius Neurons.” *Development (Cambridge, England)* 148 (11). https://doi.org/10.1242/dev.199409.

“Cell Ontology in Bioportal.” n.d. https://bioportal.bioontology.org/ontologies/CL.

“Cell Ontology in Ontology Lookup Service (OLS).” n.d. Accessed November 14, 2025. https://www.ebi.ac.uk/ols4/ontologies/cl.

Chae, Shinhyeok, Tae Joo Park, and Taejoon Kwon. 2023. “Convergent Differentiation of Multiciliated Cells.” *Scientific Reports* 13 (1): 23028.

Chen, Sijie, Yanting Luo, Haoxiang Gao, et al. 2022. “hECA: The Cell-Centric Assembly of a Cell Atlas.” *iScience* 25 (5): 104318.

Collin, Shaun P., and Ann E. O. Trezise. 2004. “The Origins of Colour Vision in Vertebrates.” *Clinical & Experimental Optometry* 87 (4-5): 217–223.

Costa, Marta, Simon Reeve, Gary Grumbling, and David Osumi-Sutherland. 2013. “The Drosophila Anatomy Ontology.” *Journal of Biomedical Semantics* 4 (1): 32.

Court, Robert, Marta Costa, Clare Pilgrim, et al. 2023. “Virtual Fly Brain—An Interactive Atlas of the Drosophila Nervous System.” *Frontiers in Physiology* 14 (January). https://doi.org/10.3389/fphys.2023.1076533.

CZI Single-Cell Biology Program, Shibla Abdulla, Brian Aevermann, et al. 2023. “CZ CELL×GENE Discover: A Single-Cell Data Platform for Scalable Exploration, Analysis and Modeling of Aggregated Data.” In *bioRxiv*. November 2. https://doi.org/10.1101/2023.10.30.563174.

Davis, Paul, Magdalena Zarowiecki, Valerio Arnaboldi, et al. 2022. “WormBase in 2022-Data, Processes, and Tools for Analyzing Caenorhabditis Elegans.” *Genetics* 220 (4). https://doi.org/10.1093/genetics/iyac003.

Deegan née Clark, Jennifer I., Emily C. Dimmer, and Christopher J. Mungall. 2010. “Formalization of Taxon-Based Constraints to Detect Inconsistencies in Annotation and Ontology Development.” *BMC Bioinformatics* 11 (October): 530.

Diehl, Alexander D., Alison Deckhut Augustine, Judith A. Blake, et al. 2011. “Hematopoietic Cell Types: Prototype for a Revised Cell Ontology.” *Journal of Biomedical Informatics* 44 (1): 75–79.

Diehl, Alexander D., Terrence F. Meehan, Yvonne M. Bradford, et al. 2016. “The Cell Ontology 2016: Enhanced Content, Modularization, and Ontology Interoperability.” *Journal of Biomedical Semantics* 7 (1): 44.

Ding, Jiarui, Xian Adiconis, Sean K. Simmons, et al. 2020. “Systematic Comparison of Single-Cell and Single-Nucleus RNA-Sequencing Methods.” *Nature Biotechnology* 38 (6): 737–746.

Domínguez Conde, C., C. Xu, L. B. Jarvis, et al. 2022. “Cross-Tissue Immune Cell Analysis Reveals Tissue-Specific Features in Humans.” *Science* 376 (6594): eabl5197.

Ebbo, Mikaël, Adeline Crinier, Frédéric Vély, and Eric Vivier. 2017. “Innate Lymphoid Cells: Major Players in Inflammatory Diseases.” *Nature Reviews. Immunology* 17 (11): 665–678.

Elmentaite, Rasa, Natsuhiko Kumasaka, Kenny Roberts, et al. 2021. “Cells of the Human Intestinal Tract Mapped across Space and Time.” *Nature* 597 (7875): 250–255.

Ergen, Can, Galen Xing, Chenling Xu, et al. 2024. “Consensus Prediction of Cell Type Labels in Single-Cell Data with popV.” *Nature Genetics*, ahead of print, November 20. https://doi.org/10.1038/s41588-024-01993-3.

Evangelista, John Erol, Zhuorui Xie, Giacomo B. Marino, Nhi Nguyen, Daniel J. B. Clarke, and Avi Ma’ayan. 2023. “Enrichr-KG: Bridging Enrichment Analysis across Multiple Libraries.” *Nucleic Acids Research*, ahead of print, May 11. https://doi.org/10.1093/nar/gkad393.

Fischer, David S., Leander Dony, Martin König, et al. 2021. “Sfaira Accelerates Data and Model Reuse in Single Cell Genomics.” *Genome Biology* 22 (1): 248.

Fischer, Felix, David S. Fischer, Roman Mukhin, et al. 2024. “scTab: Scaling Cross-Tissue Single-Cell Annotation Models.” *Nature Communications* 15 (1): 6611.

Franchini, Melania, Simona Pellecchia, Gaetano Viscido, and Gennaro Gambardella. 2023. “Single-Cell Gene Set Enrichment Analysis and Transfer Learning for Functional Annotation of scRNA-Seq Data.” *NAR Genomics and Bioinformatics* 5 (1): lqad024.

Gargano, Michael A., Nicolas Matentzoglu, Ben Coleman, et al. 2024. “The Human Phenotype Ontology in 2024: Phenotypes around the World.” *Nucleic Acids Research* 52 (D1): D1333–D1346.

Gene Ontology Consortium, Suzi A. Aleksander, James Balhoff, et al. 2023. “The Gene Ontology Knowledgebase in 2023.” *Genetics* 224 (1). https://doi.org/10.1093/genetics/iyad031.

George, Nancy, Silvie Fexova, Alfonso Munoz Fuentes, et al. 2024. “Expression Atlas Update: Insights from Sequencing Data at Both Bulk and Single Cell Level.” *Nucleic Acids Research* 52 (D1): D107–D114.

Gerbe, François, Catherine Legraverend, and Philippe Jay. 2012. “The Intestinal Epithelium Tuft Cells: Specification and Function.” *Cellular and Molecular Life Sciences: CMLS* 69 (17): 2907.

Gray, G. Kenneth, Eric G. Carlson, Tatyana Lev, et al. 2025. “Defining Breast Epithelial Cell Types in the Single-Cell Era.” *Developmental Cell* 60 (17): 2218–2236.

Hahn, Joshua, Aboozar Monavarfeshani, Mu Qiao, et al. 2023. “Evolution of Neuronal Cell Classes and Types in the Vertebrate Retina.” *Nature* 624 (7991): 415–424.

Hao, Yuhan, Stephanie Hao, Erica Andersen-Nissen, et al. 2021. “Integrated Analysis of Multimodal Single-Cell Data.” *Cell* 184 (13): 3573–3587.e29.

Hegde, Harshad, Jennifer Vendetti, Damien Goutte-Gattat, et al. 2024. “A Change Language for Ontologies and Knowledge Graphs.” September 20. http://arxiv.org/abs/2409.13906.

Hickey, John W., Elizabeth K. Neumann, Andrea J. Radtke, et al. 2022. “Spatial Mapping of Protein Composition and Tissue Organization: A Primer for Multiplexed Antibody-Based Imaging.” *Nature Methods* 19 (3): 284–295.

Hou, Wenpin, and Zhicheng Ji. 2024. “Assessing GPT-4 for Cell Type Annotation in Single-Cell RNA-Seq Analysis.” *Nature Methods* 21 (8): 1462–1465.

Hoyt, Charles Tapley, Meghan Balk, Tiffany J. Callahan, et al. 2022. “Unifying the Identification of Biomedical Entities with the Bioregistry.” *Scientific Data* 9 (November): 714.

Hoyt, Charles Tapley, Amelia L. Hoyt, and Benjamin M. Gyori. 2023. “Prediction and Curation of Missing Biomedical Identifier Mappings with Biomappings.” *Bioinformatics (Oxford, England)* 39 (4). https://doi.org/10.1093/bioinformatics/btad130.

HuBMAP Consortium. 2019. “The Human Body at Cellular Resolution: The NIH Human Biomolecular Atlas Program.” *Nature* 574 (7777): 187–192.

Hu, Congxue, Tengyue Li, Yingqi Xu, et al. 2023. “CellMarker 2.0: An Updated Database of Manually Curated Cell Markers in Human/mouse and Web Tools Based on scRNA-Seq Data.” *Nucleic Acids Research* 51 (D1): D870–D876.

Hu, Joyce, Beverly Peng, Ajith V. Pankajam, et al. 2025. “Benchmarking Single Cell Transcriptome Matching Methods for Incremental Growth of Reference Atlases.” In *bioRxiv*. April 16. https://doi.org/10.1101/2025.04.10.648034.

Hu, Yanhui, Aram Comjean, Helen Attrill, et al. 2023. “PANGEA: A New Gene Set Enrichment Tool for Drosophila and Common Research Organisms.” *Nucleic Acids Research* 51 (W1): W419–W426.

Jackson, Rebecca C., James P. Balhoff, Eric Douglass, Nomi L. Harris, Christopher J. Mungall, and James A. Overton. 2019. “ROBOT: A Tool for Automating Ontology Workflows.” *BMC Bioinformatics* 20 (1): 407.

Jackson, Rebecca, Nicolas Matentzoglu, James A. Overton, et al. 2021. “OBO Foundry in 2021: Operationalizing Open Data Principles to Evaluate Ontologies.” *Database: The Journal of Biological Databases and Curation* 2021 (October). https://doi.org/10.1093/database/baab069.

Johansen, Nelson J., Yuanyuan Fu, Matthew Schmitz, et al. 2025. “Cross-Species Consensus Atlas of the Primate Basal Ganglia.” In *bioRxivorg*. BioRxiv, December 16. https://doi.org/10.64898/2025.12.15.694496.

Jorstad, Nikolas L., Jennie Close, Nelson Johansen, et al. 2023. “Transcriptomic Cytoarchitecture Reveals Principles of Human Neocortex Organization.” *Science* 382 (6667): eadf6812.

Jupp, Simon, Tony Burdett, Catherine Leroy, and Helen E. Parkinson. 2015. “A New Ontology Lookup Service at EMBL-EBI.” *SWAT4LS* 2: 118–119.

Keefe, Matthew G., and Tomasz J. Nowakowski. 2019. “A Recipe Book for Cell Types in the Human Brain.” *Nature* 573 (7772): 36–37.

Kir, Huseyin, Ismail Ugur Bayindir, and David Osumi-Sutherland. 2024. *Ontology Based Application Starter Kit (OBASK)*. Version 0.0.1. https://doi.org/10.5281/zenodo.10518990.

Kiselev, Vladimir Yu, Tallulah S. Andrews, and Martin Hemberg. 2019. “Challenges in Unsupervised Clustering of Single-Cell RNA-Seq Data.” *Nature Reviews. Genetics* 20 (5): 273–282.

Kozareva, Velina, Caroline Martin, Tomas Osorno, et al. 2021. “A Transcriptomic Atlas of Mouse Cerebellar Cortex Comprehensively Defines Cell Types.” *Nature* 598 (7879): 214–219.

Kulmanov, Maxat, Fatima Zohra Smaili, Xin Gao, and Robert Hoehndorf. 2021. “Semantic Similarity and Machine Learning with Ontologies.” *Briefings in Bioinformatics* 22 (4). https://doi.org/10.1093/bib/bbaa199.

Li, Jin, Jun Wang, Ignacio L. Ibarra, et al. 2023. “Integrated Multi-Omics Single Cell Atlas of the Human Retina.” In *Molecular Biology*, No. Biorxiv;2023.11.07.566105v1. BioRxiv, November 8. https://www.biorxiv.org/content/10.1101/2023.11.07.566105v1.

Liu, Angela, Beverly Peng, Ajith V. Pankajam, et al. 2024. “Discovery of Optimal Cell Type Classification Marker Genes from Single Cell RNA Sequencing Data.” *BMC Methods* 1 (1). https://doi.org/10.1186/s44330-024-00015-2.

Lubiana, Tiago, Paola Roncaglia, Christopher J. Mungall, et al. 2022. “Guidelines for Reporting Cell Types: The MIRACL Standard.” In *arXiv [q-bio.OT]*. April 18. arXiv. http://arxiv.org/abs/2204.09673.

Luecken, Malte D., and Fabian J. Theis. 2019. “Current Best Practices in Single-Cell RNA-Seq Analysis: A Tutorial.” *Molecular Systems Biology* 15 (6): e8746.

Masci, Anna Maria, Cecilia N. Arighi, Alexander D. Diehl, et al. 2009. “An Improved Ontological Representation of Dendritic Cells as a Paradigm for All Cell Types.” *BMC Bioinformatics* 10 (1): 70.

Masland, Richard H. 2012. “The Neuronal Organization of the Retina.” *Neuron* 76 (2): 266–280.

Matentzoglu, Nicolas, James P. Balhoff, Susan M. Bello, et al. 2022. “A Simple Standard for Sharing Ontological Mappings (SSSOM).” *Database: The Journal of Biological Databases and Curation* 2022 (May). https://doi.org/10.1093/database/baac035.

Matentzoglu, Nicolas, Joe Flack, John Graybeal, et al. 2022. “A Simple Standard for Ontological Mappings 2022: Updates of Data Model and Outlook.” *17th International Workshop on Ontology Matching, OM 2022* 3324: 61–66.

Matentzoglu, Nicolas, Damien Goutte-Gattat, Shawn Zheng Kai Tan, et al. 2022. “Ontology Development Kit: A Toolkit for Building, Maintaining and Standardizing Biomedical Ontologies.” *Database: The Journal of Biological Databases and Curation* 2022 (October): baac087.

McLaughlin, James, Josh Lagrimas, Haider Iqbal, Helen Parkinson, and Henriette Harmse. 2025. “OLS4: A New Ontology Lookup Service for a Growing Interdisciplinary Knowledge Ecosystem.” *Bioinformatics* 41 (5): btaf279.

McMurry, Julie A., Nick Juty, Niklas Blomberg, et al. 2017. “Identifiers for the 21st Century: How to Design, Provision, and Reuse Persistent Identifiers to Maximize Utility and Impact of Life Science Data.” *PLoS Biology* 15 (6): e2001414.

Meehan, Terrence F., Anna Maria Masci, Amina Abdulla, et al. 2011. “Logical Development of the Cell Ontology.” *BMC Bioinformatics* 12 (1): 6.

Megill, Colin, Bruce Martin, Charlotte Weaver, et al. 2021. “Cellxgene: A Performant, Scalable Exploration Platform for High Dimensional Sparse Matrices.” In *bioRxiv*. April 6. https://doi.org/10.1101/2021.04.05.438318.

Miller, Corey N., Irina Proekt, Jakob von Moltke, et al. 2018. “Thymic Tuft Cells Promote an IL-4-Enriched Medulla and Shape Thymocyte Development.” *Nature* 559 (7715): 627–631.

Mungall, Christopher J., Carlo Torniai, Georgios V. Gkoutos, Suzanna E. Lewis, and Melissa A. Haendel. 2012. “Uberon, an Integrative Multi-Species Anatomy Ontology.” *Genome Biology* 13 (1): R5.

Mungall, Christopher, Alan Ruttenberg, and David Osumi-Sutherland. 2011. “Taking Shortcuts with OWL Using Safe Macros.” *Nature Precedings*, ahead of print, December 13. https://doi.org/10.1038/npre.2011.5292.2.

Munoz-Torres, Monica, and Seth Carbon. 2017. “Get GO! Retrieving GO Data Using AmiGO, QuickGO, API, Files, and Tools.” *Methods in Molecular Biology* 1446: 149–160.

Nguyen, Quy H., Nicholas Pervolarakis, Kevin Nee, and Kai Kessenbrock. 2018. “Experimental Considerations for Single-Cell RNA Sequencing Approaches.” *Frontiers in Cell and Developmental Biology* 6 (September): 108.

Nouri, Nima, Ronen Artzi, and Virginia Savova. 2025. “An Agentic AI Framework for Ingestion and Standardization of Single-Cell RNA-Seq Data Analysis.” In *bioRxiv*. August 1. https://doi.org/10.1101/2025.07.31.667880.

Noy, Natalya F., Nigam H. Shah, Patricia L. Whetzel, et al. 2009. “BioPortal: Ontologies and Integrated Data Resources at the Click of a Mouse.” *Nucleic Acids Research* 37 (Web Server issue): W170–3.

Ohlstein, Benjamin, and Allan Spradling. 2006. “The Adult Drosophila Posterior Midgut Is Maintained by Pluripotent Stem Cells.” *Nature* 439 (7075): 470–474.

Orloff, David N., Janet H. Iwasa, Maryann E. Martone, Mark H. Ellisman, and Caroline M. Kane. 2013. “The Cell: An Image Library-CCDB: A Curated Repository of Microscopy Data.” *Nucleic Acids Research* 41 (Database issue): D1241–50.

Osorno, Tomas, Stephanie Rudolph, Tri Nguyen, et al. 2022. “Candelabrum Cells Are Ubiquitous Cerebellar Cortex Interneurons with Specialized Circuit Properties.” *Nature Neuroscience* 25 (6): 702–713.

Peng, Beverly. 2025. “NS-Forest Marker Genes for Human Neocortex Cross-Areal Atlas.” Preprint, Zenodo. https://doi.org/10.5281/ZENODO.15319198.

Pranty, Abida Islam, Sara Cuevas Ocaña, and Jennifer A. Dickens. 2025. “Mechanisms of Surfactant Maturation in Development and Diseases.” *Physiology (Bethesda, Md.)*, no. physiol.00033.2025 (December). https://doi.org/10.1152/physiol.00033.2025.

Putman, Tim E., Kevin Schaper, Nicolas Matentzoglu, et al. 2024. “The Monarch Initiative in 2024: An Analytic Platform Integrating Phenotypes, Genes and Diseases across Species.” *Nucleic Acids Research* 52 (D1): D938–D949.

Regev, Aviv, Sarah A. Teichmann, Eric S. Lander, et al. 2017. “The Human Cell Atlas.” *eLife* 6 (December). https://doi.org/10.7554/eLife.27041.

Rivera, Andrea D., and David Osumi-Sutherland. 2025. “Bioinformatic Analysis of Purkinje and Molecular Layer Interneurons for WMB Annotation.” Preprint, Zenodo. https://doi.org/10.5281/ZENODO.17543831.

Sahoo, Satya S., Katja Kobow, Jianzhe Zhang, et al. 2022. “Ontology-Based Feature Engineering in Machine Learning Workflows for Heterogeneous Epilepsy Patient Records.” *Scientific Reports* 12 (1): 19430.

Scala, Federico, Dmitry Kobak, Matteo Bernabucci, et al. 2020. “Phenotypic Variation of Transcriptomic Cell Types in Mouse Motor Cortex.” *Nature*, ahead of print, November 12. https://doi.org/10.1038/s41586-020-2907-3.

Schaar, Anna Christina, Alejandro Tejada-Lapuerta, Giovanni Palla, et al. 2024. “Nicheformer: A Foundation Model for Single-Cell and Spatial Omics.” In *bioRxiv*. April 17. https://doi.org/10.1101/2024.04.15.589472.

Scheffer, Louis K., C. Shan Xu, Michal Januszewski, et al. 2020. “A Connectome and Analysis of the Adult Drosophila Central Brain.” *eLife* 9: 1–74.

Schneider, Christoph, Claire E. O’Leary, and Richard M. Locksley. 2019. “Regulation of Immune Responses by Tuft Cells.” *Nature Reviews. Immunology* 19 (9): 584–593.

Segerdell, Erik, Virgilio G. Ponferrada, Christina James-Zorn, et al. 2013. “Enhanced XAO: The Ontology of Xenopus Anatomy and Development Underpins More Accurate Annotation of Gene Expression and Queries on Xenbase.” *Journal of Biomedical Semantics* 4 (1): 31.

Shekhar, Karthik, Sylvain W. Lapan, Irene E. Whitney, et al. 2016. “Comprehensive Classification of Retinal Bipolar Neurons by Single-Cell Transcriptomics.” *Cell* 166 (5): 1308–1323.e30.

Shekhar, Karthik, and Vilas Menon. 2019. “Identification of Cell Types from Single-Cell Transcriptomic Data.” *Methods in Molecular Biology (Clifton, N.J.)* 1935: 45–77.

Shekhar, Karthik, and Joshua R. Sanes. 2021. “Generating and Using Transcriptomically Based Retinal Cell Atlases.” *Annual Review of Vision Science* 7 (September): 43–72.

Shrestha, Bindesh. 2020. “Single-Cell Metabolomics by Mass Spectrometry.” *Methods in Molecular Biology* 2064: 1–8.

Sikkema, Lisa, Ciro Ramírez-Suástegui, Daniel C. Strobl, et al. 2023. “An Integrated Cell Atlas of the Lung in Health and Disease.” *Nature Medicine* 29 (6): 1563–1577.

Sikkema, L., D. Strobl, L. Zappia, et al. 2022. “An Integrated Cell Atlas of the Human Lung in Health and Disease.” *bioRxiv*, 2022.03.10.483747.

Skarlinski, Michael D., Sam Cox, Jon M. Laurent, et al. 2024. “Language Agents Achieve Superhuman Synthesis of Scientific Knowledge.” In *arXiv [cs.CL]*. https://doi.org/10.48550/ARXIV.2409.13740.

Smith, Barry, Michael Ashburner, Cornelius Rosse, et al. 2007. “The OBO Foundry: Coordinated Evolution of Ontologies to Support Biomedical Data Integration.” *Nature Biotechnology* 25 (11): 1251–1255.

Sonder, Soren Ulrik, Matthew Plassmeyer, Denise Loizou, and Oral Alpan. 2023. “Towards Standardizing Basophil Identification by Flow Cytometry.” *Frontiers in Allergy* 4 (March): 1133378.

Spits, Hergen, David Artis, Marco Colonna, et al. 2013. “Innate Lymphoid Cells--a Proposal for Uniform Nomenclature.” *Nature Reviews. Immunology* 13 (2): 145–149.

Stefancsik, Ray, James P. Balhoff, Meghan A. Balk, et al. 2023. “The Ontology of Biological Attributes (OBA)-Computational Traits for the Life Sciences.” *Mammalian Genome: Official Journal of the International Mammalian Genome Society*, ahead of print, April 19. https://doi.org/10.1007/s00335-023-09992-1.

Subramanian, Aravind, Pablo Tamayo, Vamsi K. Mootha, et al. 2005. “Gene Set Enrichment Analysis: A Knowledge-Based Approach for Interpreting Genome-Wide Expression Profiles.” *Proceedings of the National Academy of Sciences of the United States of America* 102 (43): 15545–15550.

Sun, Xin, Anne-Karina Perl, Rongbo Li, et al. 2022. “A Census of the Lung: CellCards from LungMAP.” *Developmental Cell* 57 (1): 112–145.e2.

Tan, Shawn Zheng Kai, Shounak Baksi, Thomas Gade Bjerregaard, et al. 2025. “Digital Evolution: Novo Nordisk’s Shift to Ontology-Based Data Management.” *Journal of Biomedical Semantics* 16 (1): 6.

Tan, Shawn Zheng Kai, Huseyin Kir, Brian D. Aevermann, et al. 2023. “Brain Data Standards - A Method for Building Data-Driven Cell-Type Ontologies.” *Scientific Data* 10 (1). https://doi.org/10.1038/s41597-022-01886-2.

Tarashansky, Alexander J., Jacob M. Musser, Margarita Khariton, et al. 2021. “Mapping Single-Cell Atlases throughout Metazoa Unravels Cell Type Evolution.” *eLife* 10 (May). https://doi.org/10.7554/elife.66747.

Toro, Sabrina, Anna V. Anagnostopoulos, Susan M. Bello, et al. 2024. “Dynamic Retrieval Augmented Generation of Ontologies Using Artificial Intelligence (DRAGON-AI).” *Journal of Biomedical Semantics* 15 (1): 19.

Ualiyeva, Saltanat, Evan Lemire, Caitlin Wong, et al. 2024. “A Nasal Cell Atlas Reveals Heterogeneity of Tuft Cells and Their Role in Directing Olfactory Stem Cell Proliferation.” *Science Immunology* 9 (92): eabq4341.

Vallejo, Jenifer, Ryosuke Saigusa, Rishab Gulati, et al. 2022. “Combined Protein and Transcript Single-Cell RNA Sequencing in Human Peripheral Blood Mononuclear Cells.” *BMC Biology* 20 (1): 193.

Van Slyke, Ceri E., Yvonne M. Bradford, Monte Westerfield, and Melissa A. Haendel. 2014. “The Zebrafish Anatomy and Stage Ontologies: Representing the Anatomy and Development of Danio Rerio.” *Journal of Biomedical Semantics* 5 (1): 12.

Vasilevsky, Nicole A., Sabrina Toro, Nicolas Matentzoglu, et al. 2025. “Mondo: Integrating Disease Terminology across Communities.” *Genetics*, no. iyaf215 (October). https://doi.org/10.1093/genetics/iyaf215.

Vendetti, Jennifer, Nomi L. Harris, Michael V. Dorf, et al. 2025. “BioPortal: An Open Community Resource for Sharing, Searching, and Utilizing Biomedical Ontologies.” *Nucleic Acids Research*, ahead of print, May 13. https://doi.org/10.1093/nar/gkaf402.

Wang, Hanchen, Yichun He, Paula P. Coelho, et al. 2025. “SpatialAgent: An Autonomous AI Agent for Spatial Biology.” In *bioRxiv*. April 6. https://doi.org/10.1101/2025.04.03.646459.

Wang, Sheng, Angela Oliveira Pisco, Aaron McGeever, et al. 2021. “Leveraging the Cell Ontology to Classify Unseen Cell Types.” *Nature Communications* 12 (1): 5556.

Wang, Ye, Bin Liu, Gexin Zhao, et al. 2023. “Spatial Transcriptomics: Technologies, Applications and Experimental Considerations.” *Genomics* 115 (5): 110671.

Wilkinson, Mark D., Michel Dumontier, Ijsbrand Jan Aalbersberg, et al. 2016. “Comment: The FAIR Guiding Principles for Scientific Data Management and Stewardship.” *Scientific Data* 3: 1–9.

Winding, Michael, Benjamin D. Pedigo, Christopher L. Barnes, et al. 2023. “The Connectome of an Insect Brain.” *Science* 379 (6636): eadd9330.

Xu, Chuan, Martin Prete, Simone Webb, et al. 2023. “Automatic Cell-Type Harmonization and Integration across Human Cell Atlas Datasets.” *Cell* 186 (26): 5876–5891.e20.

Xu, Renjun, and Jingwen Peng. 2025. “A Comprehensive Survey of Deep Research: Systems, Methodologies, and Applications.” In *arXiv [cs.AI]*. June 14. arXiv. http://arxiv.org/abs/2506.12594.

Yao, Zizhen, Hanqing Liu, Fangming Xie, et al. 2020. “An Integrated Transcriptomic and Epigenomic Atlas of Mouse Primary Motor Cortex Cell Types.” *bioRxiv*, January 1, 2020.02.29.970558.

Yao, Zizhen, Cindy T. J. van Velthoven, Michael Kunst, et al. 2023. “A High-Resolution Transcriptomic and Spatial Atlas of Cell Types in the Whole Mouse Brain.” *Nature* 624 (7991): 317–332.

Zeng, Hongkui. 2022. “What Is a Cell Type and How to Define It?” *Cell* 185 (15): 2739–2755.

Zeng, Hongkui, and Joshua R. Sanes. 2017. “Neuronal Cell-Type Classification: Challenges, Opportunities and the Path Forward.” *Nature Reviews. Neuroscience* 18 (9): 530–546.

Zhao, Suyuan, Yizhen Luo, Ganbo Yang, Yan Zhong, Hao Zhou, and Zaiqing Nie. 2025. “SToFM: A Multi-Scale Foundation Model for Spatial Transcriptomics.” In *arXiv [q-bio.GN]*. July 23. arXiv. http://arxiv.org/abs/2507.11588.

Zhu, Guoli, Jiulong Hu, and Rongwen Xi. 2021. “The Cellular Niche for Intestinal Stem Cells: A Team Effort.” *Cell Regeneration (London, England)* 10 (1): 1.

# Supplementary Figure

**A**

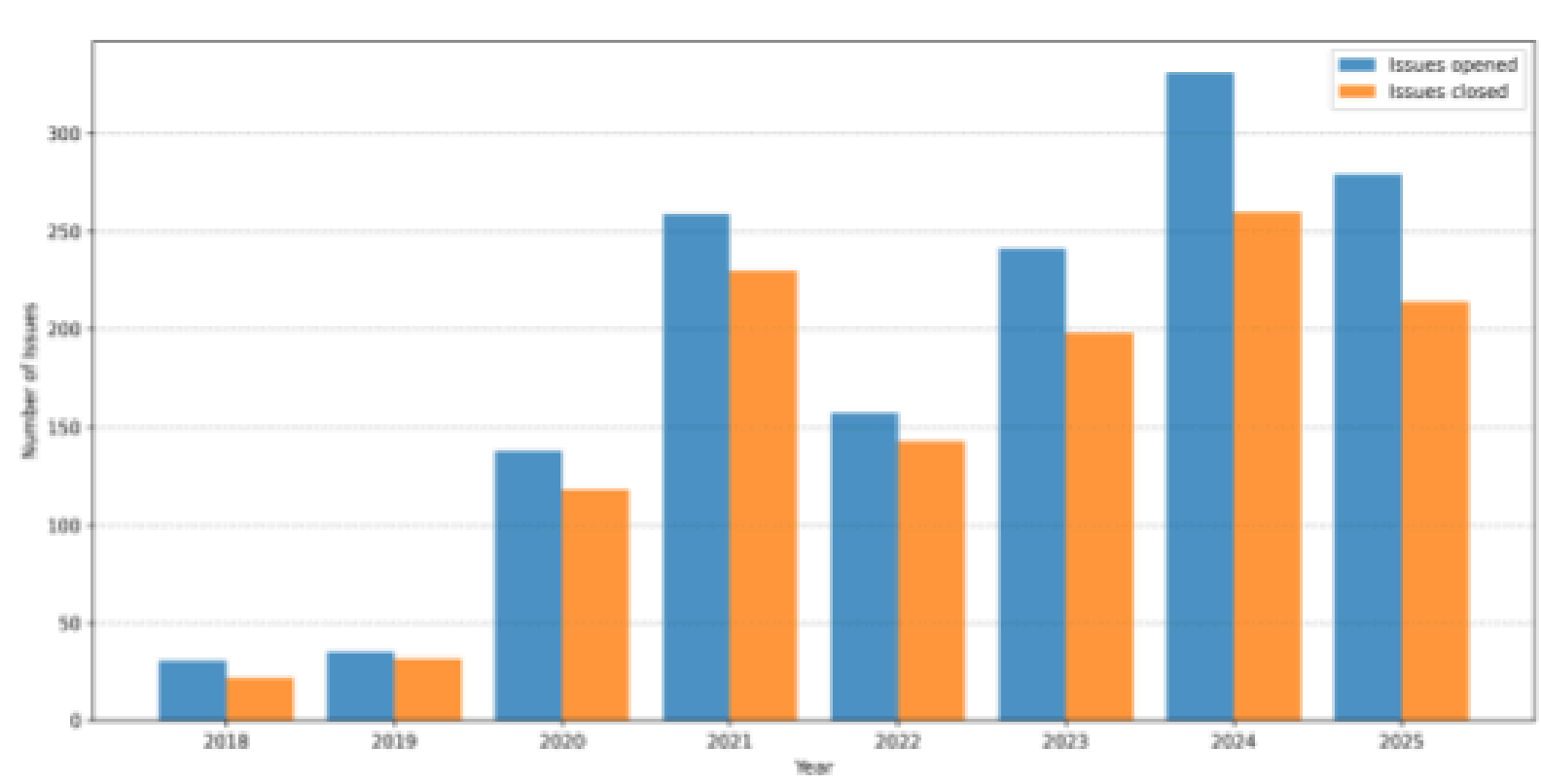

**B**

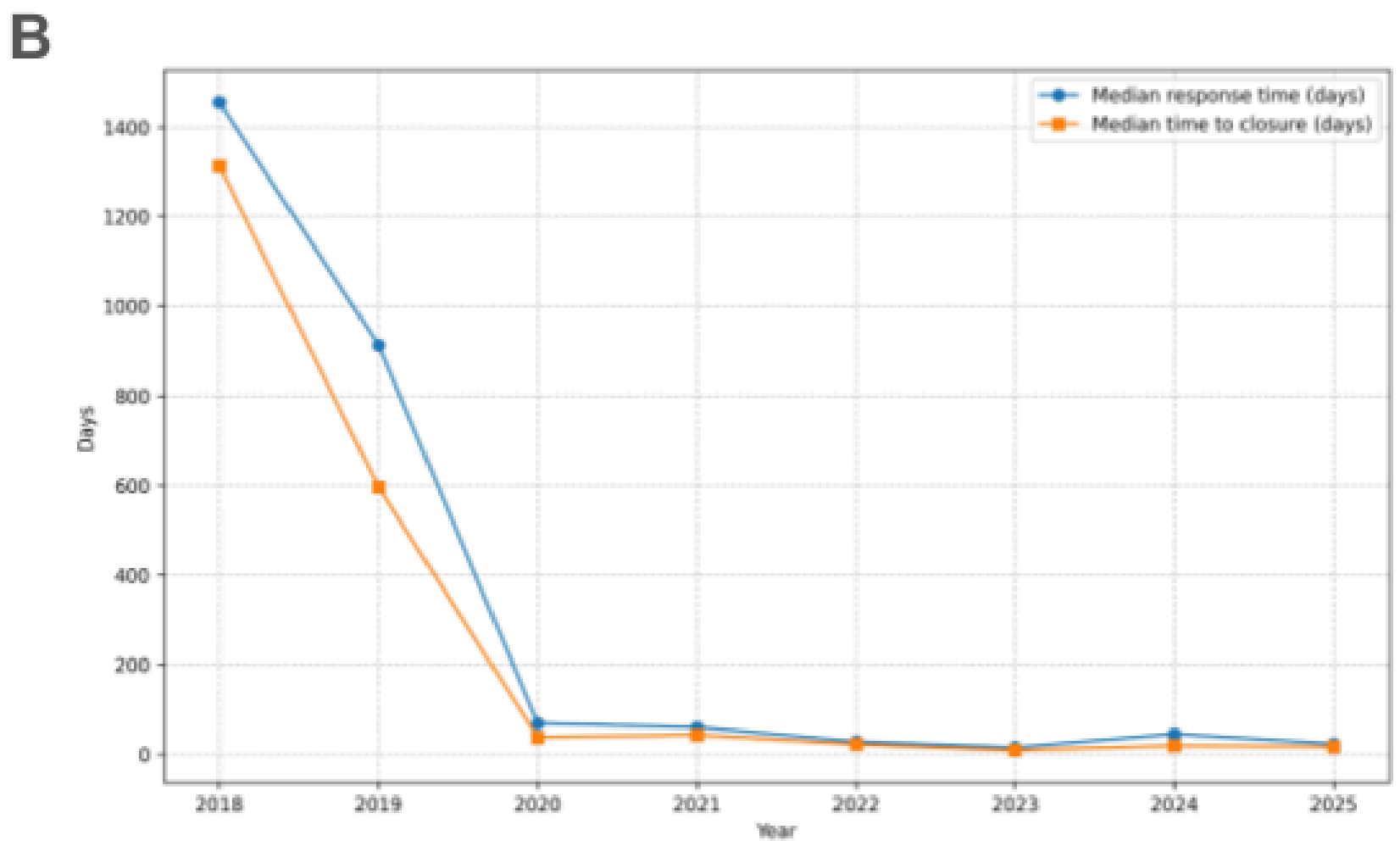

**C**

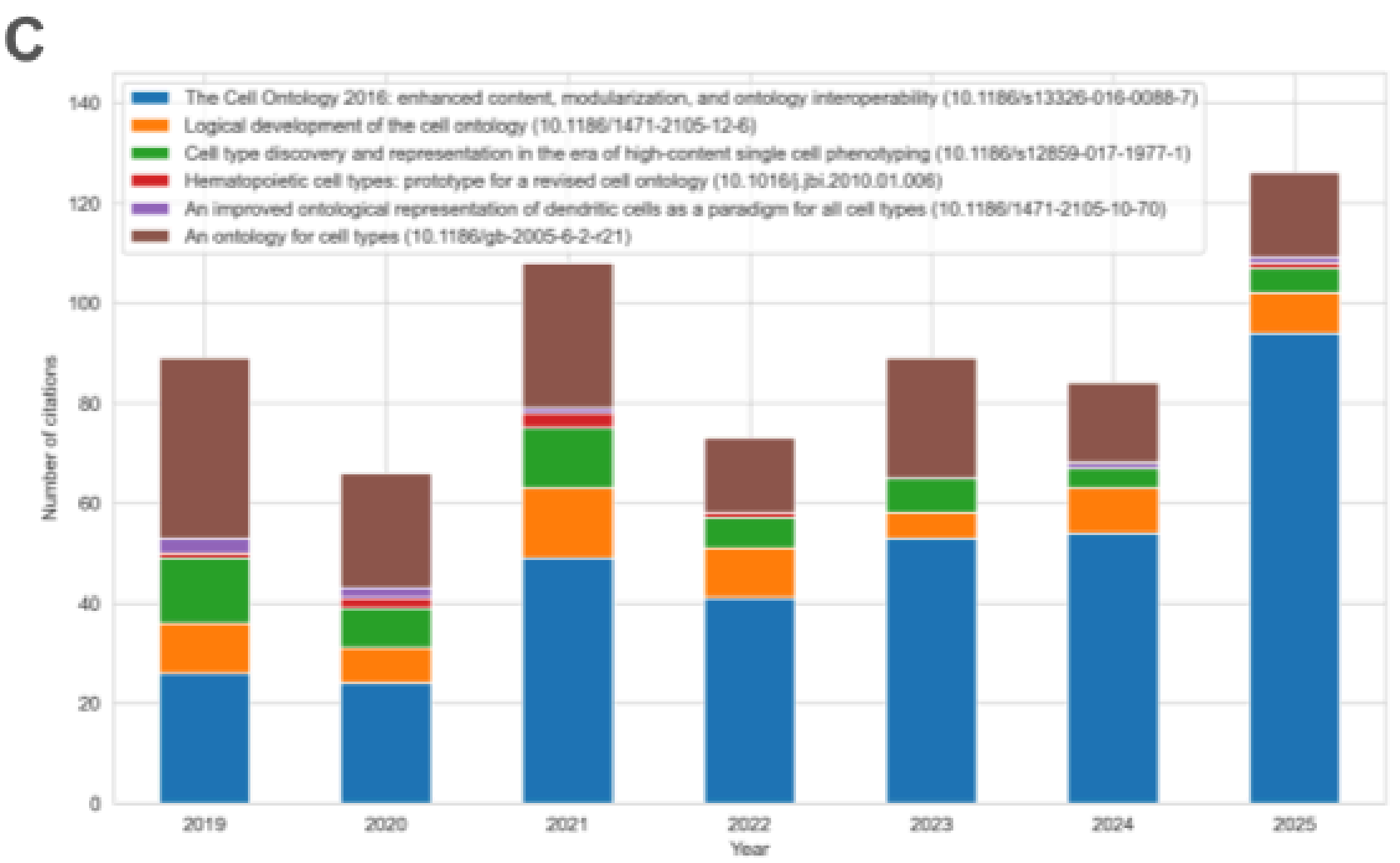